\documentclass[onecolumn,12pt,draftcls]{IEEEtran}
\usepackage{times}
\usepackage{graphicx}
\usepackage{fancyhdr}   
\usepackage{amsmath}
\usepackage{amssymb}
\usepackage{dsfont}
\usepackage{flushend}
\usepackage{bm}
\usepackage{url}
\usepackage{amsfonts}
\usepackage{dsfont}
\usepackage{amscd}
\usepackage{psfrag}
\usepackage{cases}
\usepackage{subfigure}

\usepackage{amsfonts}
\usepackage{graphicx, subfigure}
\usepackage{xcolor}
\usepackage{stfloats}
\usepackage{algorithm}
\usepackage{algpseudocode}

\begin{document}
\title{A Network-Coded Diversity Protocol for Collision Recovery in Slotted ALOHA Networks}
\author{\authorblockN{ G. Cocco$^{*}$, N. Alagha$^{\P}$, C. Ibars$^{*}$ and S. Cioni$^{\P}$}\\
\authorblockA{$^{*}$Centre Tecnol\`{o}gic de Telecomunicacions
de Catalunya -- CTTC\\ Parc Mediterrani de la Tecnologia, Av. Carl
Friedrich Gauss 7
08860, Castelldefels -- Spain\\
$^{\P}$European Space Agency - ESTEC,
Noordwijk -- The Netherlands\\
giuseppe.cocco@cttc.es, nader.alagha@esa.int,
christian.ibars@cttc.es, stefano.cioni@esa.int}\thanks{This work was
partially supported by the European Commission under project
ICT-FP7-258512 (EXALTED), by the Spanish Government under project
TEC2010-17816 (JUNTOS) and project TEC2010-21100 (SOFOCLES), and by
Generalitat de Catalunya under grant 2009-SGR-940. $^{\dag}$G. Cocco
is partially supported by the European Space Agency under the
Networking/Partnering Initiative.
Part of the present work will be presented at the 6th Advanced Satellite Multimedia Systems Conference (ASMS), to be held in Baiona (Spain) in 2012.
This work has been submitted to IEEE Journal on Selected Areas in Communications.}}
\maketitle
\abstract{\ We propose a collision recovery scheme for symbol-synchronous
slotted ALOHA (SA) based on physical layer network coding over
extended Galois Fields. Information is extracted from colliding
bursts allowing to achieve higher maximum throughput with respect to
previously proposed collision recovery schemes. An energy analysis
is also performed, and it is shown that, by adjusting the
transmission probability, high energy efficiency can be achieved.
The paper also addresses several practical aspects, namely
frequency, phase, and amplitude estimation, as well as partial
symbol asynchronism. A performance evaluation is carried out using
the proposed algorithms, revealing remarkable performance in terms
of normalized throughput.}
\begin{keywords} Physical-layer network coding; slotted ALOHA; random access networks; collision recovery; satellite communications;
CRDSA; energy efficient multiple access; spectral efficient multiple
access.
\end{keywords}
\section{Introduction}\label{sec:intro}
The throughput of Slotted ALOHA (SA) systems is limited by the
collisions that take place when more than one node accesses the
channel in the same time slot. This limitation is particularly
problematic in satellite networks with random access, where the long
round-trip time (RTT) greatly limits feedback from the receiver, for example to perform load control or to request retransmission..
Techniques like Diversity Slotted ALOHA (DSA) \cite{Choudhury_83_DSA}, in which each packet
is transmitted more than once, have been proposed in order to increase the
probability of successful detection. The spectral efficiency of SA
systems can be increased by exploiting the collided signals. In
Contention Resolution Diversity Slotted ALOHA (CRDSA)
\cite{casini07_CRDSA_aloha} the collided signals are exploited using an
iterative interference cancelation (IC) process. In CRDSA each
packet is transmitted more than once and uncollided packets are
subtracted from slots in which their replicas are present. In
\cite{bui_2010_phy_nc_aloha} a packet-level forward error correction
(FEC) code has been applied to CRDSA, while in \cite{Liva_11_CRDSA} a
convergence analysis and optimization of CRDSA has been proposed.

Another technique that allows to extract information from colliding
signals is physical layer network coding (PHY NC). PHY NC was
originally proposed to increase spectral efficiency in two-way relay
communication \cite{zhang2006_phy_nc} by having the relay decoding
the collision of two signals under the hypothesis of symbol,
frequency and phase synchronism. Several studies have been reported
in the literature about synchronization issues, gain
analysis and ad-hoc modulation techniques for PHY NC in the case of two
colliding signals
\cite{rossetto09_design_asyn_PHY_NC}\cite{louie10_practica_PHY_NC}\cite{sorensen09_PHY_NC_FSK}.
In \cite{rossetto_asms_2010} PHY NC has been applied in the satellite
context for pairwise node communication. In
\cite{durvy07_reliable_broadcast} and \cite{foh10_collision_codes}
it has been proposed to apply PHY NC to determine the identity of
transmitting nodes in case of ACK collision in multicast networks by
using energy detection and ad-hoc coding schemes, under the
hypothesis of phase synchronous signal superposition at the
receiver. In \cite{nazer11_reliable_phy_nc} the decoding of multiple
colliding signals over generally complex channels has been studied
from an information theoretical point of view. In
\cite{cocco11_mu_phy_nc_aloha} PHY NC has been applied for collision
resolution in ALOHA systems with feedback from the receiver, under
the assumption of frequency synchronous transmitters.

In this paper we present a new scheme named Network-Coded Diversity
Protocol (NCDP), that leverages on PHY NC in extended Galois Fields
for recovering collisions in symbol-synchronous SA systems. Once the
PHY NC is applied to decode the collided bursts, the receiver uses
common matrix manipulation techniques over finite fields to recover
the original messages, which results in a high-throughput scheme.
The proposed scheme and analysis differ from previous works on
collision resolutions at both system (SYS)
level and physical (PHY) level:
\begin{description}
\item[SYS:]\begin{itemize}
\item Unlike in
\cite{cocco11_mu_phy_nc_aloha}, we assume that transmissions are organized in frames. We consider two different
setups. In one, the nodes do not receive any feedback from the
receiver. If on the one hand the absence of feedback leads to a
\emph{best-effort} scheme, in which there is no guarantee for a
message to be received, on the other hand it notably simplifies the
system architecture and decreases the total amount of energy spent
per received packet. In the second setup that we consider, instead, feedback
is allowed from the receiver. In particular, we consider an
automatic repeat request (ARQ) scheme, in which a node receives an
acknowledgement (ACK) or a negative acknowledgement (NACK) from the
receiver in case a message is or is not correctly received,
respectively. A message for which a NACK has been received is
retransmitted in a different frame. The retransmission process goes
on until the message is acknowledged.
\item We evaluate jointly the
spectral efficiency (average number of messages successfully
received per slot) and the energy consumption (average amount of
energy needed for a message to be correctly received) of the
proposed scheme and compare it with other collision resolution
schemes previously proposed in the literature.
\end{itemize}
\item[PHY:] \begin{itemize}
\item We use extended Galois Fields, i.e.,
$GF(2^n)$ with $n>2$, instead of $GF(2)$,
which is generally used in PHY NC. This
allows to better exploit the diversity of
the system, leading to increased
spectral efficiency and, depending on the
system load, to an increased energy
efficiency.
\item We take into account frequency and
phase offsets at the transmitters when applying PHY NC for an arbitrary
number of colliding signals. Up to our knowledge, the issue of
frequency offsets in PHY NC has been previously addressed only for
the case of two colliding signals. See, e.g.,
\cite{maduike09_phy_nc_offsets}, \cite{lulu12_asynchronous_phy_nc}
and references therein.
\item We show the feasibility of channel
estimation for PHY NC in the presence of more than two colliding
signals, unlike previous works where only two colliding signals were
considered (see, e.g., \cite{jain11_param_estim_phy_nc}).
\item We study the effect of non perfect
symbol synchronism on the decoder FER for an arbitrary number of colliding
signals and propose four different methods
to compensate for such effect.
\end{itemize}
\end{description}

The rest of the paper is organized as follows. In Section \ref{sec:sysmod} we present the system model.
Section \ref{sec:phy-nc} describes how the channel decoding works in case of a generic number of
colliding signals with independent frequency and phase offsets. In Section \ref{sec:NCDP} the proposed
scheme is described, while its performance is studied in Section \ref{sec:protocol} in terms of both spectral and energy efficiency.
Section \ref{sec:implement} deals with issues such as channel estimation and error detection, which are fundamental for a practical implementation of the proposed scheme. Section \ref{sec:asynchro} is dedicated to the effect of imperfect symbol synchronization on the decoder performance in case of multiple colliding signals, and different schemes to overcome such effects are presented. In Section \ref{sec:num_result} we
present the numerical results, while Section \ref{sec:conclusions} contains the conclusions.
\section{System Model}\label{sec:sysmod}
Let us consider the return link (i.e, the link from a user terminal
to the satellite/base station) of a multiple access system with $M$
transmitting terminals, $T_1,.....,T_M$, and one receiver $R$.
Packet arrivals at each transmitter are modeled as a Poisson process
with rate $\frac{G}{M}$, which is independent from one transmitter
to the other. Each packet $\mathbf{u}_i=[u_i(1),....,u_i(K)]$
consists of $K$ binary symbols of information $u_i(\xi)\in \{0,1\}$,
for $ \xi=1,\ldots ,K$. We assume that, upon receiving a message,
each terminal $T_i$ uses the same linear channel code of fixed rate
$r=\frac{K}{N}$ to protect its message $\mathbf{u}_i$, obtaining the
codeword $\mathbf{x}_i=[x_i(1),...,x_i(N)]$, where $x_i(l)\in
\{0,1\}$ for $l=1,\ldots ,N$. For ease of exposition a BPSK
modulation is considered. Each codeword $\mathbf{x}_i$ is BPSK
modulated (using the mapping $0 \rightarrow -1$, $1\rightarrow +1$),
thus obtaining the transmitted signal

\begin{eqnarray}\label{eqn_tx_pam}
s_i(t)=\sum_{l=1}^Nb_i(l)g(t-lT_s),
\end{eqnarray}
where $T_s$ is the symbol period, $b_i(l)$ is the BPSK mapping of
$x_i(l)$ and $g(t)$ is the square root raised cosine (SRRC) pulse.
The signal $s_i(t)$ is called \emph{burst}.

In the following we will refer to a time division multiple access
(TDMA) scheme. However, the techniques proposed in the following can be also applied
to other access schemes, such as multi-frequency-TDMA (MF-TDMA), in
which a frame may include several carriers, or code division
multiple access (CDMA), where NCDP can be used to recover collisions
in each of the code sub-channels. It should be noted that the proposed technique still relies on single carrier transmission of each user terminal. From the user terminal perspective no significant change is required. Transmissions are organized in
frames. Each frame is divided into $S$ time slots. The number $S$ of
time slots that compose a frame is constant, i.e., it does not
change from one frame to the other. The duration of each slot is
equal to about $N$ burst symbols. When more than one burst is
transmitted in the same slot a collision occurs at the receiver. A
collision involving $k$ transmitters is said to have size $k$. We
assume symbol-synchronous transmissions, i.e., in case of a
collision, the signals from the transmitters add up with symbol
synchronism at $R$. The received signal before matched filtering and
sampling at $R$, in case of a collision of size $k$ (assuming,
without loss of generality, the first $k$ terminals collide), is:
\begin{eqnarray}\label{eqn:received1}
y(t)=h_1(t)s_1(t)+...+h_k(t)s_{k}(t)+w(t),
\end{eqnarray}
where $s_i(t)$ is the burst transmitted by user $i$, $w(t)$ is a
complex additive white Gaussian noise (AWGN) process while $h_i(t)$
takes into account the channel from terminal $i$ to the receiver.
$h_i(t)$ can be expressed as:
\begin{eqnarray}\label{eqn:chan_coeff}
h_i(t)=A_ie^{j(2\pi\Delta \nu_i t+\varphi_i)},
\end{eqnarray}
where $A_i=|h_i|$ is a lognormally distributed random variable
modeling the channel amplitude of transmitter $i$, while $\Delta
\nu_i$ and $\varphi_i$ are the frequency and phase offsets with
respect to the local oscillator in $R$, respectively. We assume that
the amplitude $A_i$ and the frequency offset $\Delta \nu_i$ remain constant
within one frame \cite{casini07_CRDSA_aloha} while $\varphi_i$ is a
random variable uniformly distributed in $[-\pi,+\pi]$ that changes
independently from one slot to the other. The fact that $\varphi_i$
changes from one slot to the other is due to the phase noise at the
transmitting terminals \cite{casini07_CRDSA_aloha}. Assuming that the
frequency offset is small compared to the symbol rate $1/T_s$
($\Delta \nu T_s\ll1$), the sample taken at time $t_l$ after matched
filtering of signal $y(t)$ is:
\begin{eqnarray}\label{eqn:received_filtered_sysmod}
r(t_l)=h_1(t_l)q_1(t_l)+...+h_k(t_l)q_k(t_l)+n(t_l),
\end{eqnarray}
where $q(t)=s(t)\oplus g(-t)$, while $n(t_l)$'s are i.i.d. zero mean
complex Gaussian random variables with variance $N_0$ in each
component. Note that even in case a BPSK modulation is used, as we
are assuming in this paper, both the I and Q components of the
received signal are considered by the receiver. This is because the
phases of the users have random relative offsets and thus both
components carry information relative to the useful signal. The
random relative offsets must be taken into account by the decoder, as
they cannot be eliminated by the demodulator. We consider this more
in detail in Section \ref{sec:phy-nc}.

We assume that the receiver has knowledge of the nodes that are
transmitting, as well as the full channel state information at each
time slot. As we are considering a random access scheme, the
knowledge about nodes identity cannot be available \emph{a priori} at
the receiver. Instead, nodes identity must be determined by $R$
starting from the received signal, even in case a collision occurs.
This can be achieved by having the transmitting nodes adding an
orthogonal preamble in each transmitted burst, assuming that the
probability that two nodes use the same preamble is negligible
\cite{casini07_CRDSA_aloha}. We discuss the issue of node
identification and channel estimation more in detail in Section
\ref{sec:implement}.
\section{Multi-User Physical Layer Network Coding}\label{sec:phy-nc}
In this section we describe the way the received signal is processed
by the receiver $R$ in case of a collision.

When a collision of size $k$ occurs, i.e., $k$ bursts collide in the
same slot, the receiver tries to decode the bit-wise XOR of the $k$
transmitted messages. This can be done by feeding the decoder with
the log-likelihood ratios (LLR) for the received signal. The
calculation of the LLRs for a collision of generic size $k$ in case
of BPSK modulation was presented in \cite{cocco11_mu_phy_nc_aloha}.
In the following we include the effect of frequency offset in the calculation of the LLRs, which was not taken into
account in \cite{cocco11_mu_phy_nc_aloha}.

When signals from $k$ transmitters collide, the
received signal at $R$ is given by (\ref{eqn:received1}). Each
codeword $\mathbf{x}_i$ is calculated from $\mathbf{u}_i$ as
$\mathbf{x}_i=\mathcal{C}(\mathbf{u}_i)$, where $\mathcal{C}(.)$ is
the channel encoder operator. All nodes use the same linear code
$\mathcal{C}(.)$.
Starting from $r(t)$, the receiver $R$ wants to decode codeword $\mathbf{x}_s\triangleq\mathbf{x}_1\oplus \mathbf{x}_2 \oplus \ldots
\oplus \mathbf{x}_k$, where $\oplus$ denotes the bit-wise XOR. In
order to do this the decoder of $R$ is fed with vector
$\mathbf{L}^\oplus=[L^\oplus(1),...,L^\oplus(N)]$ of LLRs for
$\mathbf{x}_{s}$, where:
\begin{eqnarray}\label{eqn:llrasymmetric}
L^{\oplus}(l)=\ln \left\{
\frac{\sum_{i=1}^{\lfloor\frac{k+1}{2}\rfloor} \sum_{m=1}^{{k
\choose 2i-1}} e^{-
\frac{\left|r(t_l)-\mathbf{d}^o(2i-1,m)^T\mathbf{h}(t_l)\right|^2}{2N_0}
}}{\sum_{i=1}^{\lfloor\frac{k+1}{2}\rfloor} \sum_{m=1}^{{k \choose
2i}} e^{-
\frac{\left|r(t_l)-\mathbf{d}^e(2i,m)^T\mathbf{h}(t_l)\right|^2}{2N_0}
}}\right\},
\end{eqnarray}
$\mathbf{h}(t_l)$ being a column vector containing the channel
coefficients of the $k$ transmitters at time $t_l$ (which change at
each sample due to frequency offsets), while
$\mathbf{d}^{o}(2i-1,m)$ and $\mathbf{d}^{e}(2i,m)$ are column
vectors containing one (the m-th) of the ${k \choose 2i-1}$ or ${k
\choose 2i}$ possible permutations over $k$ symbols (without
repetitions) of an odd or even number of symbols with value
``$+1$'', respectively. Equation (\ref{eqn:llrasymmetric}) is
derived considering that an even or an odd number of symbols with
value $+1$ adding up at $R$ must be interpreted by the decoder as a
0 or a 1, respectively. The derivation of $L^{\oplus}(l)$ is
detailed in the Appendix (see \cite{rossetto09_design_asyn_PHY_NC} and
\cite{sorensen09_PHY_NC_FSK} for an extension to higher order
modulations). If the decoding process is successful, $R$
obtains the message $\mathbf{u}_s\triangleq\mathbf{u}_1\oplus \ldots
\oplus \mathbf{u}_k$. In Section \ref{sec:implement} the FER curves for
different collision sizes obtained using these LLR values
are shown.
\section{Network Coded Diversity Protocol}\label{sec:NCDP}
In this section we present our network-coded diversity protocol
(NCDP) which aims at increasing the throughput and reducing packet
losses in Slotted ALOHA multiple access systems. In the first
part of the section we recall some basics of finite field
arithmetics, while in the second part we describe the NCDP at the
transmitter and at the receiver side.
\subsection{Basics of Finite Fields}\label{sec:basics_gf} A finite field is a closed set
with respect to sum and multiplication with finitely many elements.
Finite fields are often denoted as $GF(s^n)$, where $s$ is a prime
number, $n$ is a positive integer and $GF$ stands for \emph{Galois
Field}. If $n=1$ all operations (sum, subtraction, multiplication
and division) in the field coincide with operations over natural
numbers modulo $s$. If $n>1$ the field is said to be an
\emph{extended Galois Field} (EGF). In an EGF each element can be
represented as a polynomial of degree lower than $n$ and
coefficients in $GF(s)$. An element in an EGF can be represented
using the coefficients of the corresponding polynomial
representation. Thus, a string of $n$ bits can be interpreted as an
element in $GF(2^n)$. Along the same line, a string of $N=n\cdot L$
bits, $L\in \mathcal{N}$, can be represented as a vector in an
$L$-dimensional space over $GF(2^n)$ (see
\cite{koetter01_algebraic_NC} for more details).

The sum operation in an EGF is done coefficient-wise. The sum of two
elements in $GF(2^n)$ can be calculated as the bit-wise XOR of the
two $n$-bits strings corresponding to the two elements to add.

The product in an EGF can be calculated through polynomial
multiplication modulo an irreducible polynomial which characterizes
the field.
Subtraction and division are defined as the inverse operations of
sum and product, respectively, and calculated accordingly.

Finally, let us consider a system of linear equations in $GF(2^n)$
with $N^{tx}$ variables and $S$ equations, $S\geq N^{tx}$, with an
associated $S\times N^{tx}$ coefficient matrix $\mathbf{A}$ having
elements in $GF(2^n)$. The system admits a unique solution iff the
associated coefficient matrix $\mathbf{A}$ has exactly $N^{tx}$
linearly independent columns (rows).

\subsection{NCDP: Transmitter Side}\label{sec:tx_side}
\begin{figure}[!ht]
\centerline{\includegraphics[width=3.8in]{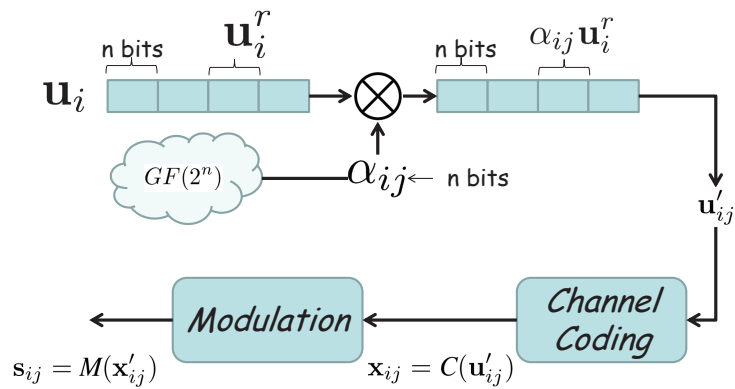}}
\caption{NCDP pre-encoding, channel coding and modulation scheme at
the transmitter side. The message to be transmitted is divided into
sub-blocks. Each sub-block is multiplied by a coefficient $\alpha_{ij}\in GF(2^n)$. Coefficients $\alpha_{ij}$, $j\in\{1,\ldots,S\}$ are chosen at random in each time slot. After the multiplication, the message is
channel-encoded, a header is attached and the modulation takes
place. } \label{fig:tx_side}
\end{figure}

Assume that node $i$ has a message $\mathbf{u}_i$ to deliver to $R$
during frame $f$. We call
\emph{active terminals} the nodes that have packets to transmit in a
given frame. Each message is transmitted more than once within a frame, i.e., several replicas of the same message are transmitted. We will give details about the number of replicas transmitted within a frame in next
section. Before each transmission, node $i$ pre-encodes
$\mathbf{u}_i$ as depicted in Fig. \ref{fig:tx_side}.
The pre-coding process works as follows. $\mathbf{u}_i$ is divided
into $L=\frac{K}{n}$ blocks of $n$ bits each. At each transmission a
different coefficient $\alpha_{ij}$,$j\in \{1,\ldots,S\}$, is drown
randomly according to a uniform distribution in $GF(2^n)$. If $\alpha_{ij}=0$,
terminal $T_i$ does not transmit in slot $j$. Each of the $L$ blocks $\mathbf{u}^r_i$,
$r\in\{1,\ldots,L\}$, is interpreted as an element in $GF(2^n)$ and
multiplied by $\alpha_{ij}$. We call $\mathbf{u}_{ij}'$ the message
$\mathbf{u}_i$ after the multiplication by $\alpha_{ij}$.
$\mathbf{u}_{ij}'$ is then channel encoded, generating the codeword
$\mathbf{x}_{ij}=C(\mathbf{u}_{ij}')$. After channel coding, a
header $p_i$ is added to $\mathbf{x}_{ij}$. Such header is chosen
within a set of orthogonal codeword (e.g. Walsh-Hadamard). The same
header $p_i$ is used for all transmissions of node $i$ within frame
$f$, i.e., it does not change within a frame. Once the header is attached, $\mathbf{x}_{ij}$ is BPSK
modulated and transmitted.

The choice of the coefficients and of the header is done as follows.
Node $i$ draws a random number $\mu$. $\mu$ is used to feed a
pseudo-random number generator, which is the same for all terminals
and is known at $R$. The first $S$ outputs of the generator  are
used as coefficients. The header is uniquely determined by $\mu$,
i.e, there is a one-to-one correspondence between the set of values
that can be assumed by $\mu$ and the set of available orthogonal
headers. The orthogonality of the preambles allows the receiver to
know which of the active terminals in frame $f$ is transmitting in
each time slot. Moreover, as the header univocally determines $\mu$
and thus the set of coefficients used by each node, $R$ is able to
know which coefficient is used by each transmitter in each slot. As
we we will see in Section \ref{sec:rx_side}, this is of fundamental
importance for the decoding process. As said before, the set of
headers is a set of orthogonal words, such as those usually adopted
in CDMA. The fundamental difference with respect to a CDMA system is
that in such system the orthogonality of the codes is used to
orthogonalize the channels and expand the spectrum, while in NCDP the
orthogonality of the preamble is used only for determining the identity of the
transmitting node, which is obtained without any spectral expansion,
as the symbol rate $1/T_s$ is equal to the chip rate (i.e., the rate
at which the modulated symbols are transmitted over the channel)
\cite{casini07_CRDSA_aloha}.
\subsection{NCDP: Receiver Side}\label{sec:rx_side}
\begin{figure}%
\centering
\parbox{3in}{%
\includegraphics[width=3.0in]{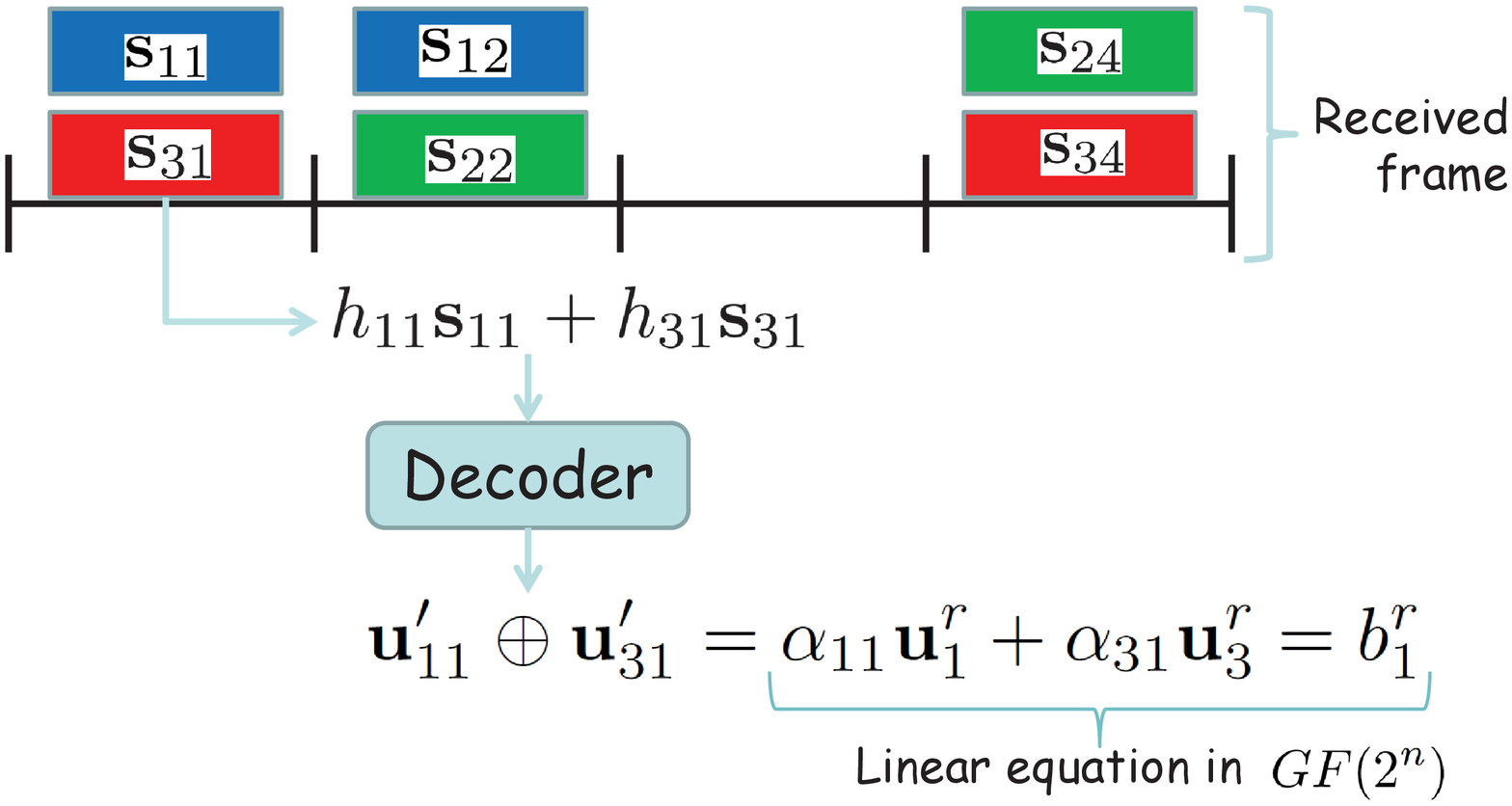}
\caption{For each of the slots the receiver uses the orthogonal preambles to determine the which node is transmitting.
With the same preamble the channel from each of the transmitters in the slot to $R$ is estimated. The channel $h_{ij}, j\in\{1,\ldots,S\}$ changes at each slot due to phase noise, according to the channel model described in Section \ref{sec:sysmod}. Once the channel has been estimated,
the decoder applies MU PHY NC to calculate the bitwise XOR of transmitted messages. The bitwise XOR corresponds to a linear equation in $GF(2^n)$
with coefficients $\alpha_{ij}$ which are known to the receiver through the header. In the figure only bursts with non-zero coefficients are shown.}
\label{fig:rx_scheme_Part1}}%
\qquad
\begin{minipage}{3in}%
\includegraphics[width=3.0in]{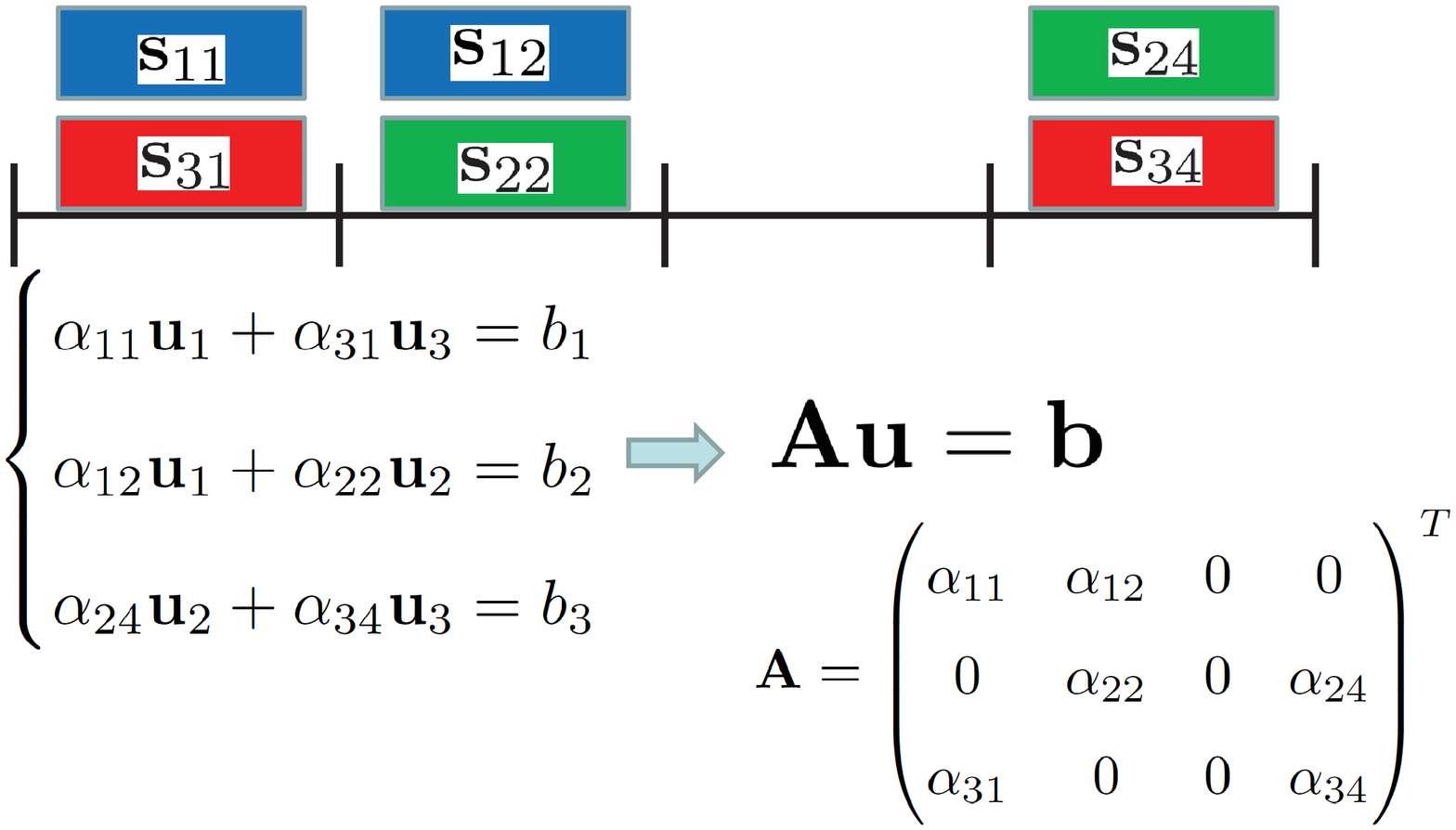}
\caption{The receiver tries to channel-decode all of the occupied
slots, thus obtaining a system of equations in $GF(2^n)$. At this
point, if the matrix $\mathbf{A}$ of coefficients is full rank, $R$
can obtain all the original messages. If $\mathbf{A}$ is not
invertible, $R$ can decode the ``clean" bursts (i.e., the bursts that
did not experience collision), then subtract them from the slots
where their replicas are. The procedure goes on until there are no more clean bursts. In the figure, $^T$ represents the transpose operator.}%
\label{fig:rx_scheme_Part2}%
\end{minipage}%
\end{figure}

The decoding scheme at the receiver side is illustrated with an
example in Fig. \ref{fig:rx_scheme_Part1} and Fig.
\ref{fig:rx_scheme_Part2}. In the example, a frame with $S=4$ slots
and $N^{tx}=3$ active terminals are considered. In each slot the
receiver uses the orthogonal preamble of each burst to determine
which node is transmitting and which coefficient has been used for
that burst. As described in Section \ref{sec:tx_side}, the
coefficients used by a node in each burst are univocally determined
by the preamble. The preamble can be determined at $R$ using a bank
of correlators which calculates in parallel the correlation of the
received signal with each element in the set of available preambles.
The preamble is also used by $R$ to estimate the channel for each of
the transmitters. The details about the channel estimation are given
in Section \ref{sec:channel_estimation}. Once the channel has been
estimated, the decoder applies PHY NC to calculate the bitwise XOR
of transmitted messages, as detailed in Section \ref{sec:phy-nc}.
The receiver tries to channel-decode the received signals using PHY
NC. According to what is stated in Section \ref{sec:basics_gf} and
Section \ref{sec:tx_side}, the bitwise XOR is interpreted as a sum
in $GF(2^n)$. Thus the slots that have been correctly decoded are
interpreted as a system of equations in $GF(2^n)$ with coefficients
$\alpha_{ij}$, which are known to the receiver through the headers
(see Fig. \ref{fig:rx_scheme_Part1}). At this point, if the
coefficient matrix $\mathbf{A}$ has full rank, $R$ can recover all
the original messages using common matrix manipulation techniques in
$GF(2^n)$ (see Fig. \ref{fig:rx_scheme_Part2}). If $\mathbf{A}$ is
not full rank, not all the transmitted packets can be recovered.
However, a part of them can still be retrieved using matrix
manipulation techniques such as Gaussian elimination. The decoding
process in case of rank deficient coefficient matrix is analyzed in
Section \ref{sec:protocol}.
\section{Throughput and Energy Analysis}\label{sec:protocol}
During each frame users buffer packets to be transmitted in the
following frame. Each node transmits its packet more than once
within a frame, randomly choosing a new coefficient in $GF(2^n)$
independently at each transmission. As described in the previous
section, the coefficients can be generated using a pseudo-random
number generator fed with a seed which is univocally determined by
the chosen orthogonal preamble. Using the preamble the receiver can
build up a coefficient matrix $\mathbf{A}$ for each frame , with
$\mathbf{A}_{i,j}=\alpha_{ij}$, $\alpha_{ij}\in\{1,\ldots,2^n-1\}$, such as the one
represented in Table \ref{tab:example}.
\begin{table}[!ht] \caption{Example of access pattern for three nodes transmitting in a frame with $S=4$ slots per frame.
$\alpha_{ij}\in GF(2^n)$ is the coefficient used by node $i$ in slot
$j$. Each coefficient can assume one of $q=2^n$ possible values,
including value $0$, which corresponds to the case in which the
terminal does not transmit.}\label{tab:example}
\begin{center}
\begin{tabular}{ c  c  c  c c}
\hline
& Slot 1 &Slot 2& Slot 3 & Slot 4 \\ \hline
$T_1$ & $\alpha_{11}$      & $\alpha_{12}$      & $\alpha_{13}$        & $\alpha_{14}$         \\ \hline
$T_2$ & $\alpha_{21}$      & $\alpha_{22}$      & $\alpha_{23}$        & $\alpha_{24}$         \\ \hline
$T_3$ & $\alpha_{31}$      & $\alpha_{32}$      & $\alpha_{33}$        & $\alpha_{34}$      \\ \hline
\end{tabular}
\end{center}
\end{table}
Columns represent time slots while rows represent the active
terminals, i.e., the terminals that transmit in present frame.
If $\alpha_{ij}=0$, terminal $i$ does not transmit in slot $j$.
During time slot $j$, $R$ receives the sum of the bursts with $\alpha_{ij}\neq 0$. From the received
signal, $R$ tries to obtain the bit-wise XOR of the encoded messages
as described in Section \ref{sec:sysmod}. The XOR is interpreted by
$R$ as a linear equation in $GF(2^n)$, the coefficients of which are
derived through the orthogonal preamble as described in Section
\ref{sec:NCDP}. If $N^{tx}$ is the number of active terminals in a
frame and assuming that all the received signals are
decoded correctly, a linear system of equations in $GF(2^n)$ is
obtained with $S$ equations and $N^{tx}$ variables. Each variable
corresponds to a different source message. If $\mathbf{A}$ has rank
equal to $N^{tx}$, then all the messages can be obtained by $R$. A
necessary condition for $\mathbf{A}$ to be full rank is $N^{tx}\leq
S$, i.e., the number of active terminals in a frame must be lower than the
number of slots in a frame. Assuming Poisson arrivals with aggregate
intensity $G$, the probability of such event is:
\begin{eqnarray}\label{eqn:prob_n_greater_S}
Pr\{N^{tx}\leq S\}=\sum_{n=0}^{S}\frac{(GS)^ne^{-GS}}{n!},
\end{eqnarray}
which includes also the case in which there are no active terminals
during a frame. For instance, in case of $S=100$ slots and $G=0.8$
the probability expressed by (\ref{eqn:prob_n_greater_S}) is on the order of $0.99$. Even if $N^{tx}<S$, however, it
can still happen that $\mathbf{A}$ is not full rank, i.e., not all
the messages can be recovered. The probability that $\mathbf{A}$ is
full rank for a given $N^{tx}<S$ depends on the MAC policy, and
particularly on the probability distribution used to choose the
coefficients.

One possibility is to use a uniform distribution for the
coefficients (i.e., each coefficient can assume any value in
$\{0,\ldots,2^n-1\}$ with probability $2^{-n}$). In this case the
number $d$ of transmitted replicas is a random variable, and the probability
that $\mathbf{A}$ is full rank is \cite{trullols11_exact_NC}:
\begin{eqnarray}\label{eqn:prob_dec_NC}
P(S,N_{tx})=\prod_{k=0}^{N_{tx}-1}\left(1-\frac{1}{2^{n(S-k)}}\right).
\end{eqnarray}
Using (\ref{eqn:prob_n_greater_S}) and (\ref{eqn:prob_dec_NC}) we
find the expression for the normalized throughput:
\begin{align}\label{eqn:analytic_throughput}
\Phi&=\frac{1}{S}\sum_{m=1}^{S}m\frac{(GS)^me^{-GS}}{m!}P(S,m)&\notag\\
&=\frac{1}{S}\sum_{m=1}^{S}\frac{(GS)^me^{-GS}}{(m-1)!}\prod_{k=0}^{m-1}\left(1-\frac{1}{2^{n(S-k)}}\right)&\notag\\
&=G\sum_{m=0}^{S-1}\frac{(GS)^me^{-GS}}{m!}\prod_{k=0}^{m}\left(1-\frac{1}{2^{n(S-k)}}\right).&
\end{align}
From Eqn. (\ref{eqn:analytic_throughput}) we can see that $\Phi$
grows with $n$, which means that the system throughput increases
with the size of the considered finite field. Moreover, we have:
\begin{align}\label{eqn:asympt_optimality}
\lim_{n\rightarrow \infty}\Phi&=\lim_{n\rightarrow \infty}
\left[G\sum_{m=0}^{S-1}\frac{(GS)^me^{-GS}}{m!}\prod_{k=0}^{m}\left(1-\frac{1}{2^{n(S-k)}}\right)\right]&\notag\\
&=G\sum_{m=0}^{S-1}\frac{(GS)^me^{-GS}}{m!}.&
\end{align}
From Eqn. (\ref{eqn:asympt_optimality}) it can be seen that the
normalized throughput $\Phi$ tends to the probability of having less
than $S$ transmitters in a frame as $n\rightarrow \infty$.

The MAC scheme we just analyzed presents one main drawback in terms
of the energy efficiency of the protocol. As a matter of fact, given
the frame length $S$, a node transmits each message on average
$E[d]=S\times p$ times, $p=(1-2^{-n})$ being the probability to
choose a non-zero coefficient, i.e., the average number of transmissions
grows linearly with $S$. In order to decrease the energy
consumption, the probability of choosing the zero coefficient may be
increased. However, a reduction in the transmission probability $p$
may affect the system throughput. In order to understand the
relationship between the probability $p$ and the throughput $\Phi$,
we refer to some results in random matrix theory. The problem can be
formulated as follows: consider an $N^{tx}\times S$ random matrix
$\mathbf{A}$ over $GF(2^n)$ with i.i.d. entries, each of which
assumes value $0$ with probability $p$ while with probability $1-p$
it assumes values in $\{1,\ldots,2^n-1\}$. We are interested in the
relationship between $p$ and the probability that $\mathbf{A}$ is
full rank. In \cite{blomer97_rank_sparse_matrix} the authors show
that, if we want to achieve a rank $N^{tx}-O(1)$ with high
probability, then, for $N^{tx}$ large, $p$ cannot be lower than
$\frac{\log(N^{tx})}{N^{tx}}$. At high loads (i.e., $G\simeq 1$), on
average $N^{tx}\simeq S$, which means that, setting
$p=\frac{\log(S)}{S}$, the average number of transmissions (and so
the energy consumption) for each node is $E[d]=\log(S)$, i.e., it
grows logarithmically with the number of slots in a frame. On the
other side, $S$ must be kept large enough, as this increases the
decoding probability, which makes the choice of small $S$
unpractical. With reference to the example considered earlier in
this section, in which $S=100$, the average number of transmissions
corresponding to the minimum required $p$ is equal to about $4.6$.
\begin{figure}[!ht]
\centerline{\includegraphics[width=3.8in]{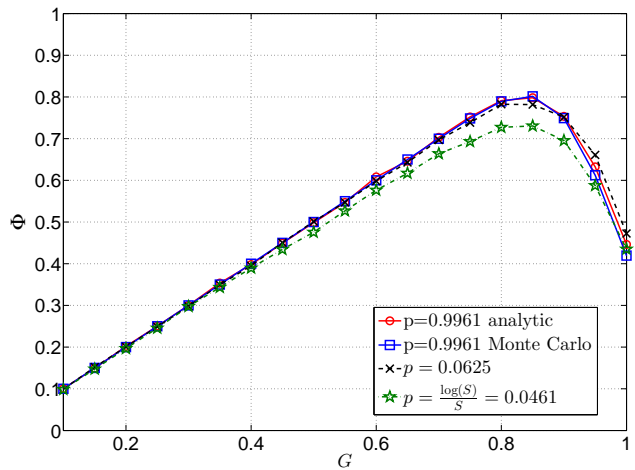}}
\caption{Normalized throughput plotted against the normalized
offered load for different values of the transmission probability
$p$. We set $S=100$ slots per frame while the coefficients were
chosen in $GF(2^8)$.} \label{fig:lower_bound_phy_nc}
\end{figure}
We evaluated numerically the effect a reduction of $p$ has on $\Phi$
for the case $S=100$ and $q=2^8$. We considered three cases. In the
first one the transmission probability in each slot has been set to
$p=1-2^{-n}=0.9961$, which corresponds to the case studied in the
first part of this section and for which the throughput is given by
Eqn. ($\ref{eqn:analytic_throughput}$). In the second case we set
$p$ just above the threshold, i.e.,
$p=0.0625>\frac{\log(S)}{S}=0.0461$, while in the last case $p$ has
been set exactly equal to the threshold probability. Fig.
\ref{fig:lower_bound_phy_nc} shows the results together with the
numerical validation of Eqn. ($\ref{eqn:analytic_throughput}$). It
is interesting to note how passing from $p=0.9961$ to $p=0.0628$,
with a reduction in transmission probability (or, equivalently, in
average energy per message) of about $93.7\%$, leaves the throughput
unchanged, while a further decrease of $p$ of just another $1.5\%$
leads to a $10\%$ reduction in the maximum throughput with respect
to the case $p=0.9961$.

To further lower the energy consumption and control the number of
repetitions $d$ (which, being a Bernoulli random variable, can
theoretically assume values as large as $S$), an alternative is to
fix the number of transmitted replicas \emph{a priori}. Although this
solution may lead in some cases to the impossibility of decoding all
the transmitted messages, it may still be possible to recover many
of them by using Gaussian elimination.
\section{Implementation Aspects}\label{sec:implement}
\subsection{Channel Estimation and Node Identification}\label{sec:channel_estimation}
For each frame the receiver $R$ needs to know which of the active
terminals is transmitting in each slot and must have channel state
information for each of the users. Both needs are addressed
including an orthogonal preamble, such as the spreading codes used
in CDMA, at the beginning of the burst. The use of an orthogonal
preamble was proposed in \cite{casini07_CRDSA_aloha} for the
estimation of the phase in collided bursts. In
\cite{casini07_CRDSA_aloha} frequency offset and channel amplitude
are derived from the clean bursts (i.e., bursts that did not
experience collisions) and assumed to remain constant over the whole
frame. Unlike in \cite{casini07_CRDSA_aloha}, the method we propose
does not rely only on clean bursts. Thus the
frequency offset and the amplitude of each transmitter must be
estimated using the collided bursts for each frame. Although the
performances of the estimator are likely to degrade with respect to
the clean burst case, especially in case of high order collisions,
the estimation can leverage in the information of all the collided
bursts, which improves the estimation. For instance, if a packet is
transmitted twice during a given frame and experiences collisions of
order 2 in the first transmission and 4 in the second, the two
estimations can be combined to obtain a better estimation of
amplitude and frequency offset, which are constant during the whole
frame.

In order to prove the feasibility of channel estimation in such
conditions we show the results we obtained using the Estimate
Maximize (EM) algorithm. We adopted the approach described in
\cite{feder88_param_estim_EM}, where the EM algorithm is used to
estimate parameters from superimposed signals. In
\cite{feder88_param_estim_EM} two examples were proposed related to
multipath delay estimation and direction of arrival estimation. We
apply the same approach to estimate amplitudes, phases and frequency
offsets from the baseband samples of the received signal in case of
a collision of size $k$. The algorithm is divided into an $E$ step,
in which each signal is estimated, and an $M$ step, in which the
mean square error between the estimation made at the $E$ step of
current iteration and the signal reconstructed using parameters
calculated in previous iteration is minimized with respect to the
parameters to estimate. Formally, once initialized the parameters
with randomly chosen values, at each iteration we have the following
two steps:

\emph{Estimation step} - for $i=1,\ldots,k$ calculate

\begin{eqnarray}\label{eqn:E_step}
&\hat{p}_i^{(n)}(t)&=b_i(t)\hat{A}_i^{(n)}e^{j(2\pi \widehat{\Delta\nu}_i^{(n)}T_s t+\hat{\varphi}_i^{(n)})} \notag\\
&+& \beta_i\left[r(t)-\sum_{l=1}^k b_l(t)\hat{A}_l^{(n)}e^{j(2\pi
\widehat{\Delta\nu}_l^{(n)}T_st+\hat{\varphi}_l^{(n)})}\right],
\end{eqnarray}

\emph{Maximization step} - for $i=1,\ldots,k$ calculate

\begin{eqnarray}\label{eqn:M_step}
\min_{A',\Delta\nu',\varphi'}\sum_{t=1}^{N^{pre}}\left|b_i(t)\hat{p}_i^{(n)}(t)-A'e^{j(2\pi
\Delta\nu' T_st + \varphi')}\right|^2,
\end{eqnarray}
where $p_i(t)$ is the preamble of burst $i$ after the matched
filter, $A'$, $\Delta\nu'$ and $\varphi'$ are tentative values for
the parameters to be estimated, $N^{pre}$ is the preamble length,
$b_i(t)\in \{\pm 1\}$ is the t-th symbol in the preamble of the i-th
node and $T_s$ is the sampling period, taken equal to the symbol
rate. $\beta_i$ are free parameters that we arbitrarily set to
$\beta_i=0.8$, for $i=1,\ldots,k$.
\begin{figure}[!ht]
\centerline{\includegraphics[width=3.8in]{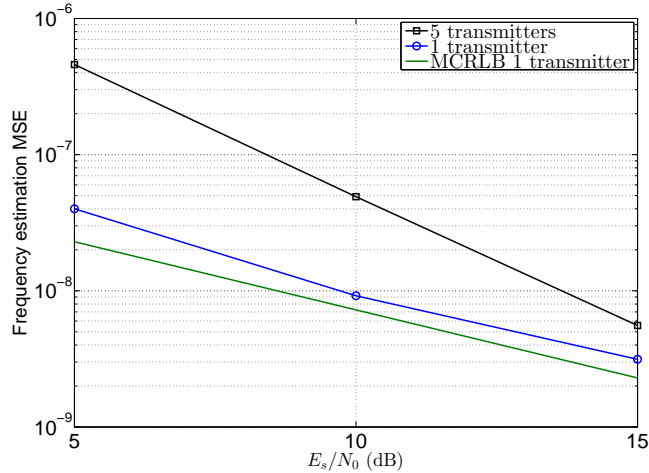}}
\caption{Mean squared error (MSE) of the frequency offset estimation, i.e.,
$E[|\widehat{\Delta\nu}-\Delta\nu|^2]$. $E_s$ is the average
energy per transmitted symbol for each node. The modified Cramer-Rao
lower bound (MCRLB) for the case of one transmitter is also shown
for comparison.} \label{fig:frequency_estimation}
\end{figure}
\begin{figure}[!ht]
\centerline{\includegraphics[width=3.8in]{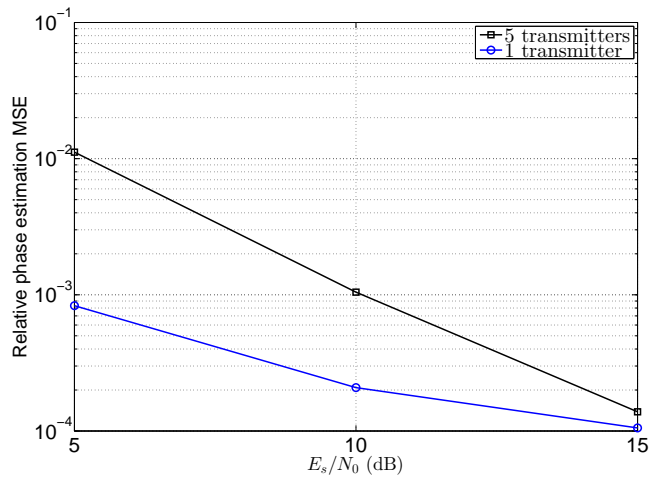}}\caption{MSE of the phase offset estimation
normalized to $\pi$, i.e., $E[|\hat{\varphi}-\varphi|^2]/\pi^2$. $E_s$
is the average energy per transmitted symbol for each node.}
\label{fig:phase_estimation}
\end{figure}
\begin{figure}[!ht]
\centerline{\includegraphics[width=3.8in]{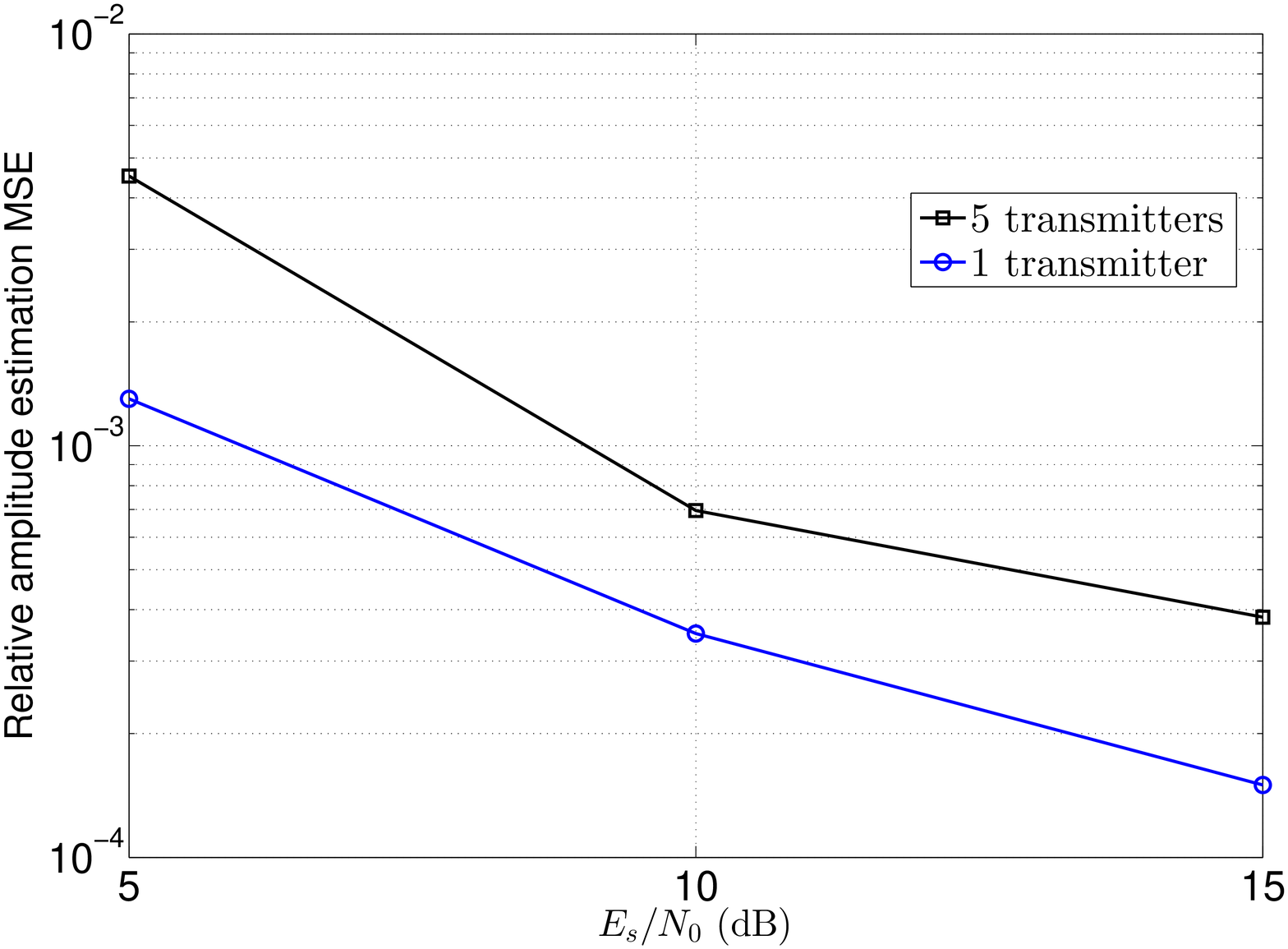}}
\caption{MSE of the amplitude estimation normalized to
the actual amplitude of the channel, i.e, $E[|\hat{A}-A|^2/A^2]$.
$E_s$ is the average energy per transmitted symbol for each node.}
\label{fig:amplitude_estimation}
\end{figure}

We evaluated numerically the performance of the EM estimator
assuming that phase offsets are uniformly distributed in
$[-\pi,+\pi]$, frequency offsets are uniformly distributed in
$[0,\Delta \nu^{max}]$ with $\Delta \nu^{max}$ equal to $1\%$ of the
symbol rate on the channel ($1/T_s$), and amplitudes are
log-normally distributed. Figures \ref{fig:frequency_estimation},
\ref{fig:phase_estimation} and \ref{fig:amplitude_estimation} show
the mean squared error (MSE) of the estimation error for
frequency, phase and amplitude, respectively. Amplitude error is
normalized to the actual amplitude value while phase
error is normalized to $\pi$. In the
simulations we used as preambles Walsh-Hadamard words of length 128
symbols. The EM algorithm was run twice starting from randomly
chosen initial values of the parameters and taking as result the
values of the parameters that lead to the minimum of the sum across
the signals of the error calculated in the last E step. This was
done in order to reduce the probability to choose a ``bad" local
maximum, which is a problem that affects all the ``hill climbing"
algorithms. For each run 6 iterations were made.

\begin{figure}[!ht]
\centerline{\includegraphics[width=4in]{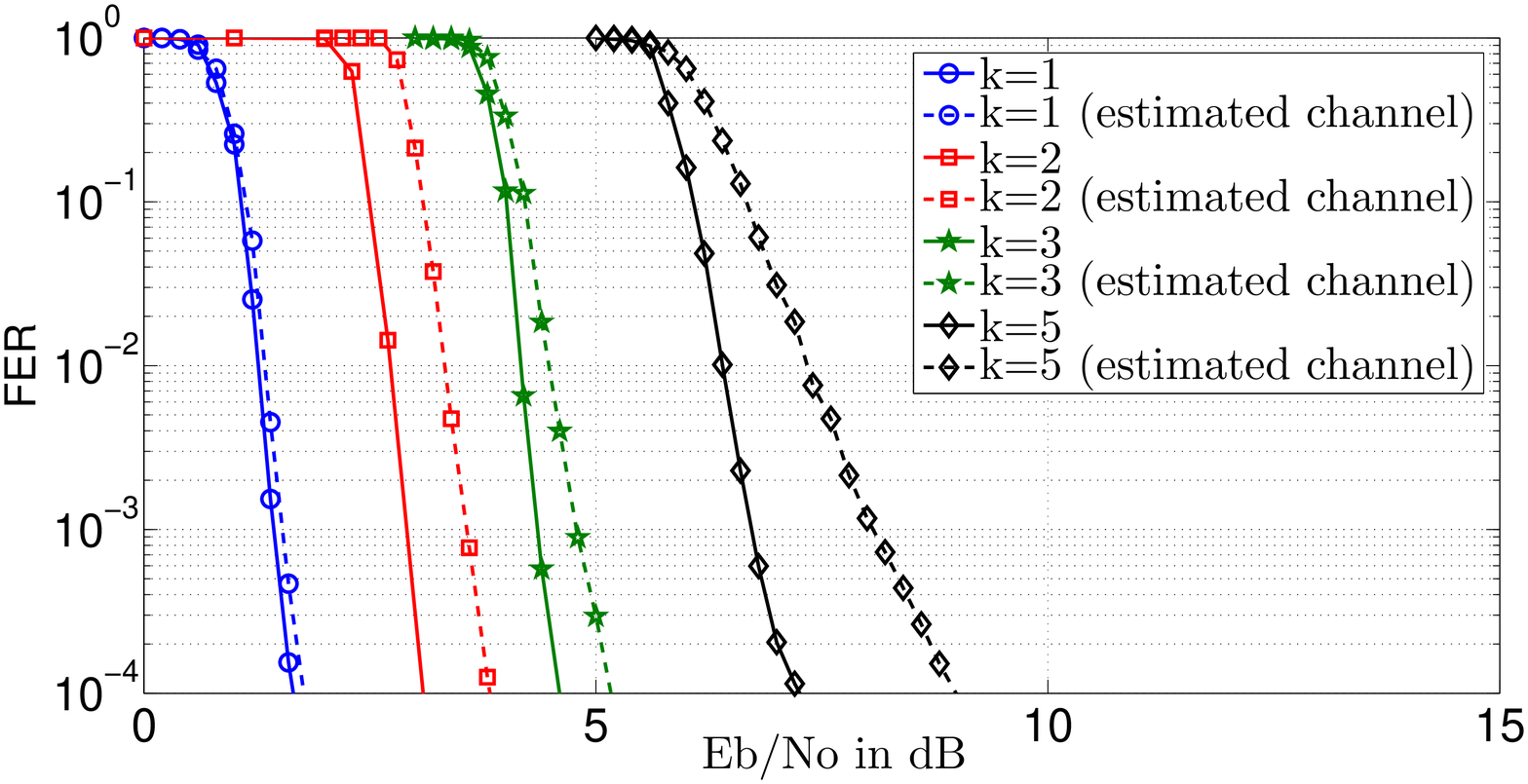}}
\caption{FER for the XOR of transmitted messages for different
numbers of transmitters. $E_b$ is the energy per information bit for
each node. A tail-biting duo-binary turbo code with rate $1/2$ and
codeword length $1504$ symbols is used by all nodes. Phase offsets
are uniformly distributed in $[-\pi,+\pi]$, frequency offsets are
uniformly distributed in $[0,\Delta \nu^{max}]$ with $\Delta
\nu^{max}$ equal to $1\%$ of the symbol rate on the channel.
Amplitudes are constant and equal to 1. The FER curves for the case of estimated channels using the EM algorithm are also shown.} \label{fig:fer}
\end{figure}

 In Fig. \ref{fig:fer} the FER curves for
different collision sizes obtained using the LLR values calculated in Section \ref{sec:phy-nc}
are shown. The plots are obtained using a tail-biting duo-binary turbo code
with rate $1/2$ and codeword length
equal to 1504 symbols. The phase offsets $\varphi_i$ are random
variables uniformly distributed in $[-\pi,+\pi]$ while frequency
offsets are uniformly distributed in $[0,\Delta \nu^{max}]$ with
$\Delta \nu^{max}$ equal to $1\%$ of the symbol rate $1/T_s$. The FER curves for the case of estimated channels using the EM algorithm are also shown.
\subsection{Error Detection}\label{sec:CRC}
An important issue in slotted ALOHA is the capability of the
receiver to determine whether the received bursts are correctly
decoded or not. This is particularly important in NCDP, where the
error made in the decoding of a collision can propagate possibly
leading to the loss of a whole frame. A common practice in packet
networks is the use of a cyclic redundancy check (CRC), which allows
to detect a wrong decoding with a certain probability. Some CRC's
are based on a field which is appended to the message before channel
coding, called \emph{CRC field}. As the CRC operations are done in
$GF(2)$ and by the linearity of the channel encoder, the CRC field
in the message obtained by decoding a collision of size $k$ is a
good CRC for $\mathbf{u}_s$, which is the bitwise XOR of the
messages encoded in the $k$ collided signals. This allows to detect
decoding errors, within the limits of the CRC capabilities, also in
collided bursts. The implementation aspect of what type of CRC
should be used is out of scope of this paragraph.
\section{Performance of Multi User Physical Layer Network Coding
with Imperfect Symbol Synchronization}\label{sec:asynchro}
 In Section \ref{sec:sysmod}
we assumed that signals from different receivers add up with symbol
synchronism at the receiver in case of a collision.
\begin{figure}[!ht]
\centerline{\includegraphics[width=4in]{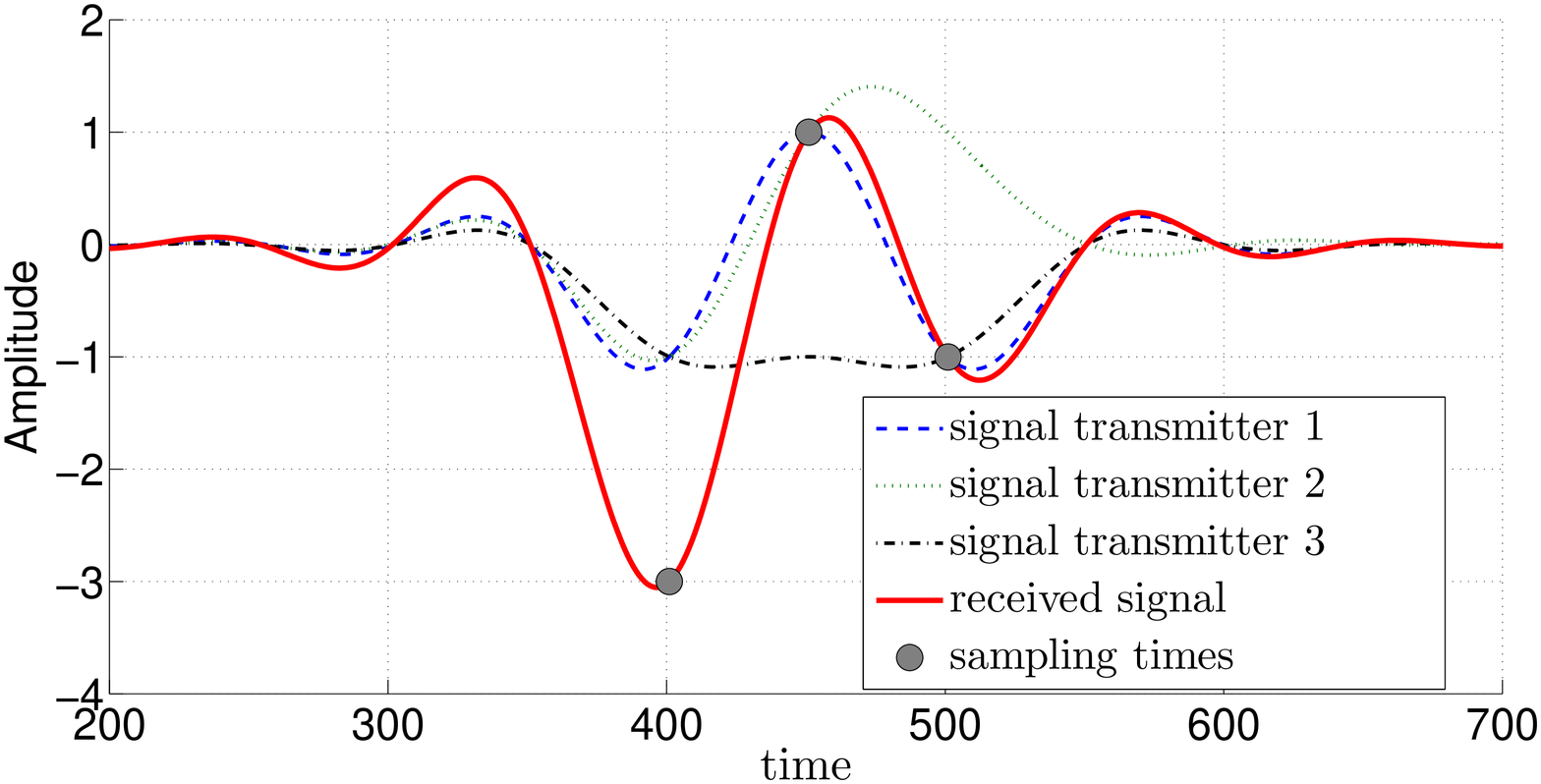}}
\caption{Received signal after the matched filter in case of three
colliding bursts with no timing offsets, i.e., $\Delta T_1=\Delta T_2=\Delta T_3=0$. The
transmitted signals after the matched filter in case of collision-free reception are also shown. The transmitted symbols are: [-1
1 -1], [-1 1 1] and [-1 -1 -1] for transmitter 1, 2 and 3,
respectively. For sake of clarity, frequency and phase offsets
as well as channel amplitudes were not included in the plot and the signals
were considered as real. The samples, shown with grey circles in the figure, are taken at instants corresponding to the optimal sampling instants for each of the signals as if they were received without experiencing collision.} \label{fig:received_aligned}
\end{figure}
In Fig. \ref{fig:received_aligned} an example is shown of received signal and sampling instants in the case of
three nodes transmitting with no timing offsets. The transmitted signals, which are also shown, modulate
the sets of symbols [-1 1 -1], [-1 1 1] and [ -1 -1 -1]. The
situation depicted in the figure is an illustrative one, as in a
real system both I and Q signal components are present, signals may have
different amplitudes, phase and frequency offsets for each of the
bursts and the signal is immersed in thermal noise. However, in a
real system there will always be a certain symbol misalignment,
which grows larger as the resources dedicated to the synchronization
phase diminish (see, e.g., \cite{nee94_cdma_synchro} and references
therein for examples of synchronization algorithms). Being able to
cope with non perfect symbol synchronism can bring important
advantages, such as less stringent constraints on signal alignment,
with consequent savings in terms of network resources needed for the
synchronization. In this section we study the effect of non perfect
symbol synchronization and propose possible countermeasures. Let us
consider a slotted multiple access with $k$ nodes accessing the
channel at the same time. We assume that each transmitter has its
own phase and frequency offsets. We further assume that each burst
falls completely within the boundaries of a time slot, i.e., no
burst can fall between two consecutive time slots. Let us call $T'$
the time at which the peak of the first symbol of the bursts that
first arrives at $R$. We define the \emph{relative delay} (RD)
$\Delta T_i$ of node $i$ as the temporal distance between the peak
value of the first pulse of burst $i$ and $T'$. In other words, the
burst which arrives first at the receiver is used as reference, i.e.,
has RD equal to $0$. We assume SRRC pulses with roll off factor
$\alpha$ are used. We further assume that all RD's belong to the
interval $[0,\Delta T^{max}]$, with $0\leq\Delta T^{max}\leq T_s/2$.

In case of a collision of $k$ bursts, the received signal before the
matched filter is:
\begin{eqnarray}\label{eqn:received2}
y(t)=\sum_{i=1}^{k}s_i(t) +w(t),
\end{eqnarray}
where,
\begin{eqnarray}\label{eqn:received3}
s_i(t)=A_i\sum_{l=1}^Nb_i(l)g(t-lT_s-\Delta T_i)e^{j(2\pi\Delta
\nu_it+\varphi_i)},
\end{eqnarray}
$N$ being the number of symbols in the burst, $g(t)$ is the square
root raised cosine pulse and $w(t)$ represents an AWGN process. The
samples taken after the matched filter at times $t_l$ are:
\begin{eqnarray}\label{eqn:received_filtered}
r(t_l)=y(t)\otimes g(-t)\mid_{t=t_l}=\sum_{i=1}^{k}q_i(t_l)+n(t_l),
\end{eqnarray}
where,
\begin{eqnarray}\label{eqn:received_filtered2}
q_i(t_l)=A_i\sum_{l=1}^Nb_i(l)p(t_l-lT_s-\Delta T_i)e^{j(2\pi\Delta
\nu_i t_l+\varphi_i)},
\end{eqnarray}
$p(t)$ being the raised cosine pulse, $\otimes$ is the convolution
operator and $n(t)$ is the noise process after filtering and
sampling. Note that in (\ref{eqn:received_filtered2}) the
exponential term is treated as a constant. This approximation is
done under the assumption that $\Delta \nu T_s\ll1$, i.e., the
exponential term is almost constant over many symbol cycles.

The sampled signal is then sent to the channel decoder. It is not
clear at this point which is the optimal sampling time, as the
optimal sampling time for each of the bursts taken singularly may be different.
Moreover, sampling the signal just once may not be the optimal
choice. Actually, as we will show in next section, the performance
of the decoder is quite poor in case a single sample per symbol is
taken.

In the following we propose several techniques to mitigate
the impairment due to imperfect symbol synchronization. We assume
that $R$ has knowledge of the relative delays of all the
transmitters, which can be derived through the orthogonal preambles.
We further assume that $R$ has perfect CSI for each of the
transmitters. Without loss of generality and for ease of exposition,
from now on we will refer to the sampling time for the symbol number
1.
\subsection{Single sample}
\paragraph{Mean Delay} The first method we present is \emph{Mean Delay}
(MD). In MD the received signal is sampled just once per symbol. The
sampling time is chosen to be the mean of the relative delay, i.e.:
\begin{eqnarray}\label{eqn:mean_delay}
T^{MD}=\frac{1}{k}\sum_{m=1}^k\Delta T_m.
\end{eqnarray}
The sample $r(T^{MD})$ is then used to calculate the LLR's as in
Eqn. (\ref{eqn:llrasymmetric}). ISI is not taken into account.
\subsection{Multiple samples}
In the following we describe four different methods that use $k$ samples
per symbol, $k$ being the collision size.

\begin{figure}[!ht]
\centerline{\includegraphics[width=4in]{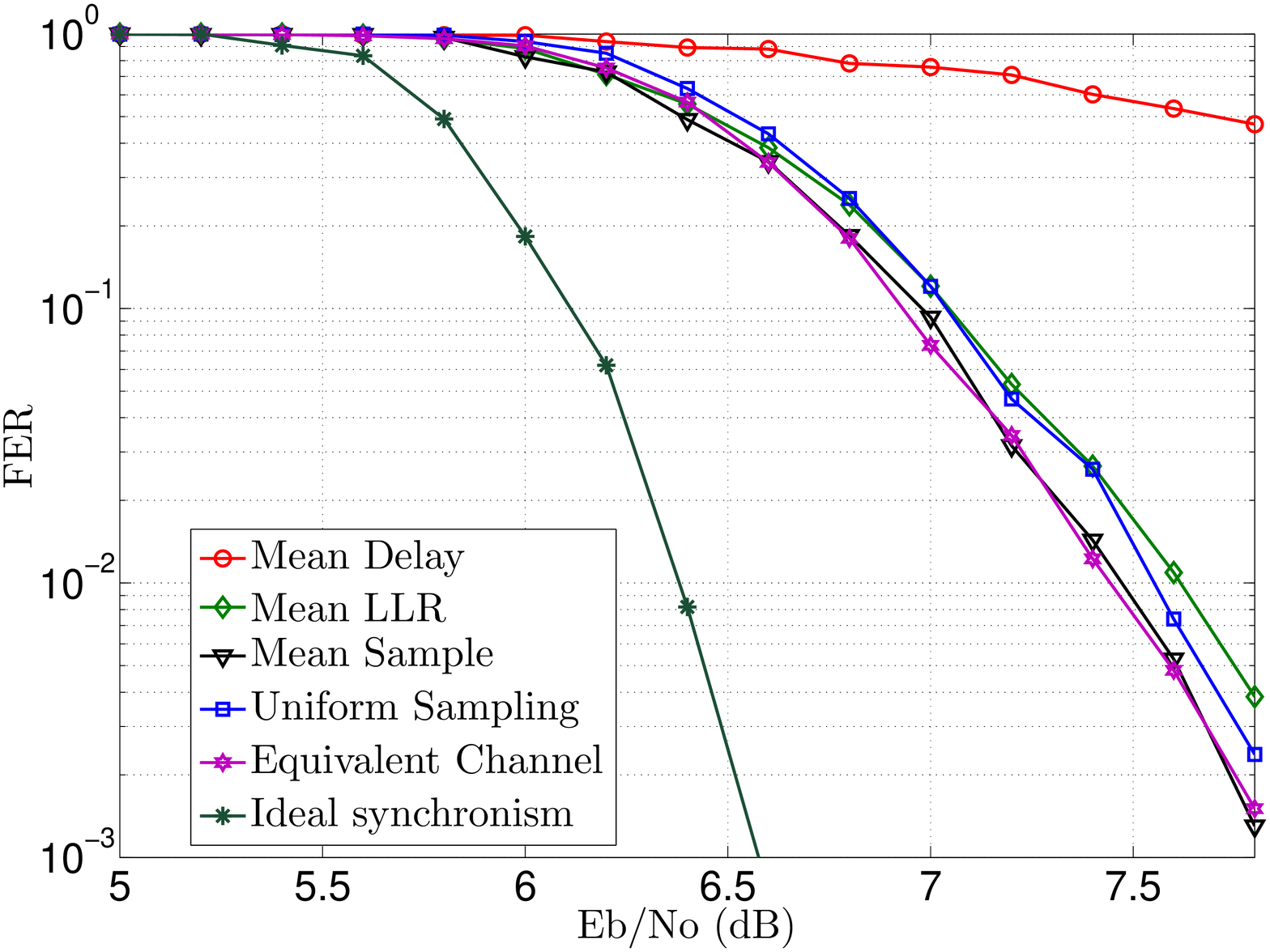}}
\caption{Frame error rate for decoding a collision of size 5 with
independent frequency and phase offsets across the transmitters and
delays uniformly distributed in $[0,T_s/4]$. A roll-off factor of $\alpha=0.35$ was used. The results for the 5
different methods are shown together with the FER for the case of
ideal symbol synchronism. Oversampling significantly improves the
FER with respect to the case of single sample. The two methods that
exploit knowledge of relative delays, i.e, MS and EC, perform
slightly better than the others. The FER of all methods present a
lower slope w.r.t. the ideal case, losing about 1 dB at
$FER=10^{-2}$ for the methods that use more than one sample.}
\label{fig:FER_compare_async_1_4}
\end{figure}

We start by describing two methods in
which the symbol is sampled $k$ times in correspondence of the RD's.
Due to the non perfect synchronization, when the signal is sampled
in $\Delta T_i$ the sample obtained is the sum of the first symbol
of each of the users, weighted by the relative channel coefficient,
plus a term of ISI due to signals $s_j, j\in\{1,\ldots,k\}, j\neq
i$, which are sampled at non ISI-free instants. As the LLR's need
the channels of each of the users, the ISI should be taken into
account. However, the ISI is a function of many (theoretically all)
symbols, and can not be taken into account exactly.
\begin{figure}[!ht]
\centerline{\includegraphics[width=4in]{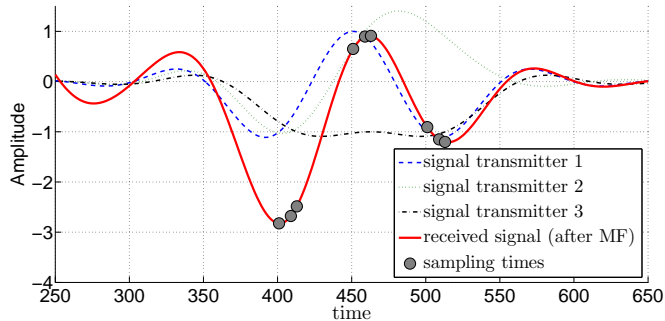}}
\caption{Received signal after the matched filter in case of three
colliding bursts with timing offsets $\Delta T_1=0, \Delta T_2=T_s/6$ and $\Delta T_3=T_s/4$. The
transmitted signals after the matched filter in the case of collision-free reception are also shown. The transmitted symbols are: [-1
1 -1], [-1 1 1] and [-1 -1 -1] for transmitter 1, 2 and 3,
respectively. The samples, shown with grey circles in the figure, are taken at instants corresponding to the optimal sampling instants for each of the signals as if they were received without experiencing collision. Unlike in the case of perfect symbol alignment, here more than one sample per symbol is taken.} \label{fig:received_misaligned}
\end{figure}
In Fig. \ref{fig:received_misaligned} the received signal after the matched filter is shown in the case of three
colliding bursts with timing offsets $\Delta T_1=0, \Delta T_2=T_s/6$ and $\Delta T_3=T_s/4$. The
transmitted signals after the matched filter in the case of collision-free reception are also shown. The symbols transmitted by each terminal are the same as in Fig. \ref{fig:received_aligned}. The samples, shown with grey circles in the figure, are taken in correspondence of the RD's, which coincide with the optimal sampling instants for each of the signals as if they were received without experiencing collision.
\paragraph{Mean LLR} In
\emph{Mean LLR} (ML) the received signal is sampled $k$ times in the
instants correspondent to $\Delta T_i$, $i=1,\ldots,k$. For each of
the samples the LLR's are calculated as in
(\ref{eqn:llrasymmetric}). Then the average of the $k$ LLR's is
passed to the decoder.
\paragraph{Mean Sample} As in ML, also in \emph{Mean
Sample} (MS) $r(t)$ is sampled $k$ times in correspondence of the
relative delays. The difference between the two methods is that in
MS the samples are averaged out to obtain the mean sample:
\begin{eqnarray}\label{eqn:average_sample}
\overline{r}(t)=\frac{1}{k}\sum_{m=1}^kr(\Delta T_m).
\end{eqnarray}Finally, $\overline{r}(t)$ is used in
the (\ref{eqn:llrasymmetric}) instead of $r(t)$.
\paragraph{Uniform
Sampling} In \emph{Uniform Sampling} (US) the signal is sampled $k$
times as in previous methods, but the sampling times do not
correspond to the RD's. The sampling times are chosen uniformly in
$[0,\Delta T^{max}]$, i.e, in case of $k$ transmitters the samples
are taken at intervals of $\Delta T^{max}/(k-1)$. Then, as in MS,
the samples are averaged out and used in the calculation of the LLRs.
This method has the advantage that receiver does not need the
knowledge of the RD's in order to decode and the sampling itself is
simplified as it is done uniformly in each symbol.
\paragraph{Equivalent Channel} The received signal is sampled $k$
times in the instants correspondent to $\Delta T_i$, $i=1,\ldots,k$.
In the method \emph{Equivalent Channel} (EC) the amplitude variation of the
channel of each user due to imperfect timing is taken into account
for the current symbol. Note that the ISI is not taken into account,
but only the variation in amplitude of present symbol due to
imperfect timing is accounted for. Assuming that the received signal
is sampled at time $t=\Delta T_i$, then the channel coefficient of
burst $q$ that is used in the LLR is:
\begin{eqnarray}\label{eqn:equiv_channel_coeff}
h^{eq}_q(t)=A_qe^{j(2\pi\Delta \nu_qT_s \Delta
T_i+\varphi_q)}p(\Delta T_i-\Delta T_q),
\end{eqnarray}
$p(t)$ being the raised cosine pulse.
After the sampling, the
$k$ samples per symbol are averaged together and used in the LLR
instead of $r(t)$.
This sampling procedure is equivalent (apart
from the ISI) to filtering the received signal using a filter which is
matched not to the single pulse, but to the pulse resulting from the
delayed sum of $M$ pulses.
In Fig. \ref{fig:FER_compare_async_1_4} the frame error rate is
shown for the case of 5 transmitters with delays uniformly
distributed in $[0,T_s/4]$. Constant channel amplitudes were
considered, while phases and frequency offsets are i.i.d. random
variables in $[0, 2\pi]$ and $[0, \Delta \nu_{max}]$ respectively,
where $\Delta \nu_{max}$ is equal to $1/(100T_s)$. The results for
the 5 different methods are shown together with the FER for the case
of ideal symbol synchronism. The methods that use more than one
sample per symbol perform significantly better than MD, which uses
only one sample per symbol. Among the methods based on oversampling,
MS and EC perform slightly better than the other two. The FER of all
methods present a lower slope w.r.t. the ideal case. The loss is
about 1 dB at $FER=10^{-2}$ for the methods that use oversampling.
\section{Numerical Results}\label{sec:num_result}
In this section we present the numerical results. Our performance
metrics are the normalized throughput $\Phi$ defined as:
\begin{eqnarray}\label{eqn:throughput}
\Phi=G(1-\Upsilon),
\end{eqnarray}
where $\Upsilon\in[0,1]$ is the average packet loss rate (i.e, the
ratio of the number of lost packets to the total number of packets
that arrive at the transmitters), and the average energy consumption
per received message $\eta$, defined as the average
number of transmissions needed for a message to be correctly
received by $R$. We consider two benchmarks. The first one is a system that
implements the contention resolution diversity slotted ALOHA (CRDSA)
protocol, which has been proposed in \cite{casini07_CRDSA_aloha}. In
CRDSA a node transmits two or more copies of a burst (twin bursts)
in different slots randomly chosen within a frame. Each of the twin
bursts contains information about the position of the other twin bursts in
the frame. If one of the twin bursts does not experience a collision
(i.e, it is \emph{clean}) and can be correctly decoded, the position
of the other twin bursts is known. These bursts may or may not
experience a collision with other bursts. If it happens, these are
removed through interference cancelation using the decoded bursts.
In order to do this $R$ memorizes the whole frame, decodes the clean
bursts, reconstructs the modulated signals and, once the effect of
each user's channel has been included in the reconstruction, they
are subtracted from the slots in which their replicas are located.
The IC process is iterated  for a number $N^{iter}$ of times, at
each time decoding the bursts that appear to be ``clean" after the
previous IC iteration. The second benchmark is a slotted ALOHA
system.

We consider two different setups. In one, the nodes do not receive
any feedback by the receiver, while in the second setup $R$ gives some feedback to the active terminals. For this last case we consider an automatic repeat request
(ARQ) scheme, in which a node receives an acknowledgement (ACK) or a
negative acknowledgement (NACK) from the receiver in case a message
is or is not correctly received, respectively. An alternative to the
NACK is to having the transmitters using a counter for each
transmitted packet, indicating the time elapsed since it has been
transmitted. If the timer exceeds a threshold value (which depends
on the system's RTT), the message is declared to be lost. A node
that receives a NACK (or whose timer exceeds the threshold vale)
enters a \emph{backlog state}. Backlogged nodes retransmit the
message for which they received the NACK in another frame, uniformly
chosen at random among the next $B$ frames. We call $B$ the
\emph{maximum backlog time}. The process goes on until the message
is acknowledged \cite{Kleinrock75_performance}. In both setups we
assume a very large population of users. Furthermore, we assume that
the average SNR is high enough so that the FER at the receiver is
negligible.

In the first setup, in which no feedback is provided by the
receiver, the average amount of energy spent by a node for each
message which is correctly received does not change with the system
load $G$, and is equal to the average number of times a message is
repeated within a frame.
\begin{figure}[!ht]
\centerline{\includegraphics[width=4.2in]{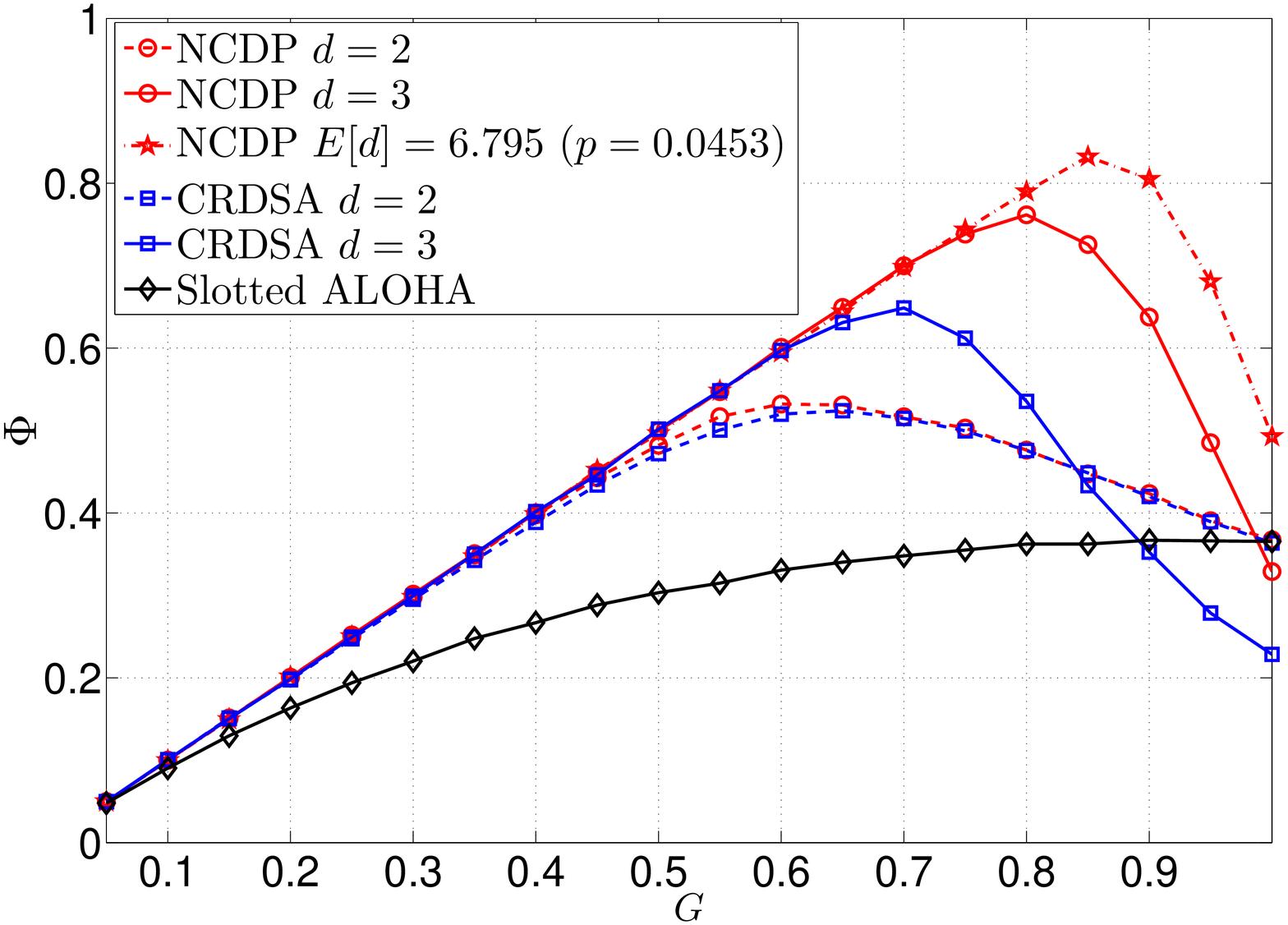}} \caption{Normalized throughput $\Phi$ vs normalized traffic load
$G$. The normalized traffic load is the average rate at which new
messages are injected in the network, and is independent from the
number of times a message is repeated within a slot. In the
simulation the frame size was set to $S=150$ slots. No feedback was
assumed from the receiver.} \label{fig:through_no_feed}
\end{figure}In Fig. \ref{fig:through_no_feed} the normalized throughput $\Phi$
is plotted against the normalized traffic load $G$. The normalized
traffic load is the average rate at which the \emph{new messages}
(i.e, messages which are being transmitted for the first time) are
injected in the network, and is independent from the number of times
a message is repeated within a slot. In the figure, the throughput
curves of NCDP and CRDSA schemes in case of $d=2$ and $d=3$ replicas
are shown. The throughput curve for NCDP in case of a constant
retransmission probability $p=0.0453$ is also shown. Note that this probability is above the threshold value we mentioned in Section \ref{sec:protocol}, as for $S=150$ we have $\log(S)/S=0.0334$. The scheme with $p=0.0453$ outperforms all the others in terms of throughput, achieving
a peak value of about $0.8$. It is interesting to note how
increasing the number of transmissions per message (and so the
energy consumption) leads to an increase in the peak throughput of
the system. However, $\Phi$ increases about $0.2$ when passing from
$d=2$ to $d=3$ repetitions, while the increase in the peak
throughput is only about $0.05$ when passing from $d=2$ repetitions
per message to an average of $E[d]=6.795$ in case of a fixed
transmission probability.

\begin{figure}[!ht]
\centerline{\includegraphics[width=4.2in]{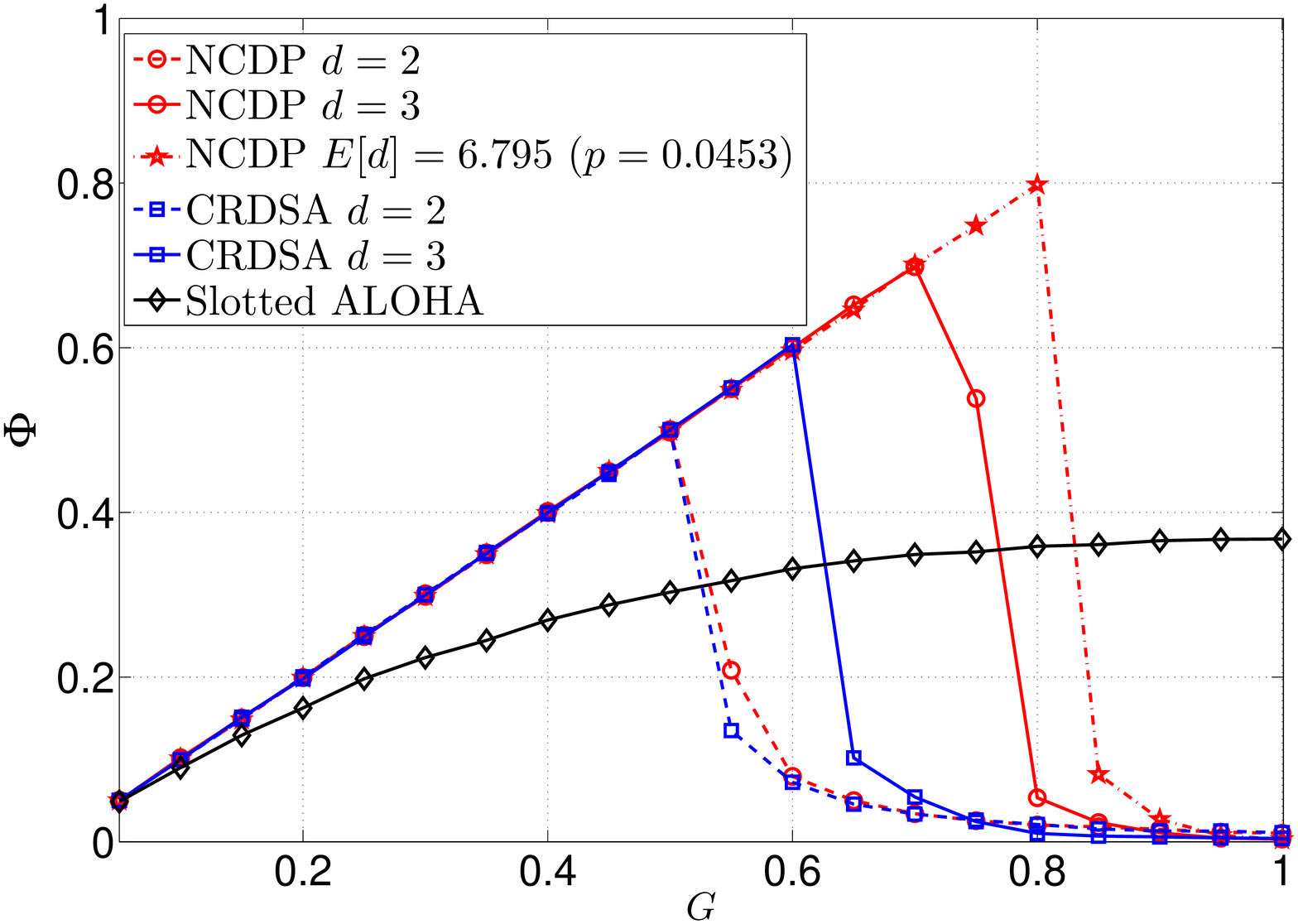}} \caption{Normalized throughput $\Phi$ vs normalized traffic load
$G$ in a system with retransmission. In the simulation the frame
size was set to $S=150$ slots while the maximum backlog time was set
to $B=50$ frames.} \label{fig:through_feed}
\end{figure}

\begin{figure}[!ht]
\centerline{\includegraphics[width=4.2in]{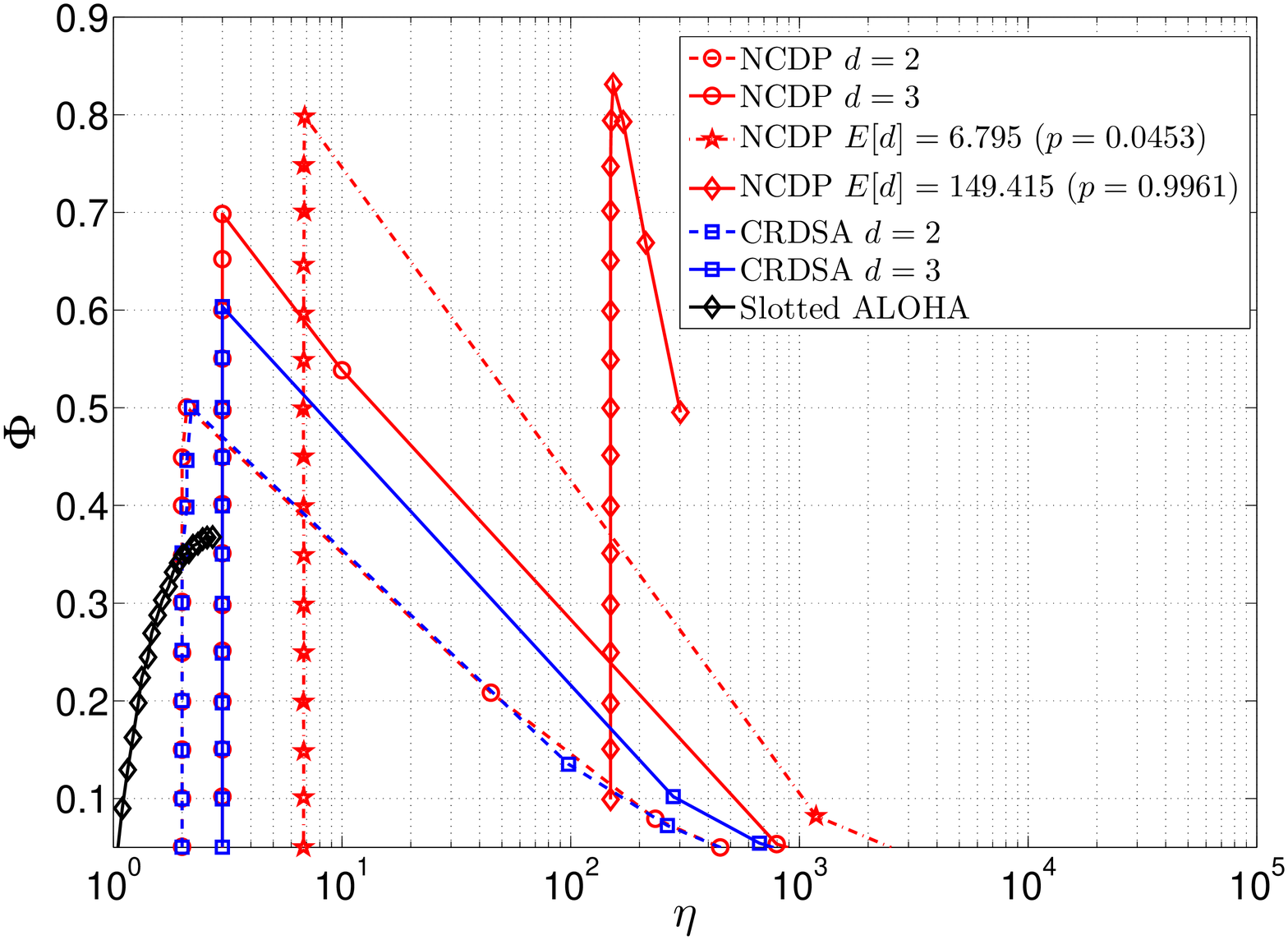}} \caption{Normalized throughput vs average energy consumption per
decoded message for $S=150$ and $B=50$ frames.}
\label{fig:through_energy}
\end{figure}

 In the second setup, in which retransmissions are allowed, we evaluate jointly the
spectral efficiency (average number of messages successfully
received per slot) and the energy consumption (average number of
transmissions needed for a message to be correctly received) of the
schemes under study. In Fig. \ref{fig:through_feed}, $\Phi$ is
plotted against $G$ for a frame size $S=150$ slots and a maximum
backlog time $B=50$ frames. The figure shows how $\Phi$ increases
linearly with $G$ up to a threshold load value. Such threshold
increases with the (average) number of repetitions of the considered
scheme. The $\Phi$ curve of NCDP upperbounds that of CRDSA. The
reason for this lies in the way the decoding process is carried out by
the receiver $R$ in NCDP. $R$ first tries to decode the whole frame,
which is feasible if the coefficient matrix
$\mathbf{A}$ has rank $N^{tx}$. If the whole frame can not be decoded, then
$R$ applies Gaussian elimination on $\mathbf{A}$, in order to
recover as many messages as possible. It can be easily
verified that Gaussian elimination in NCDP is the
equivalent, in a finite field, of the IC process of CRDSA, which is
applied in the analog domain.
In order to compare jointly the spectral and the energy efficiency
of the different schemes, we plot the curves for the normalized
throughput vs the average energy consumption per received message
$\eta$, which is shown in Fig. \ref{fig:through_energy}.
The increase in throughput coming from an increased number of
transmissions implies a higher energy consumption for a given
transmitter in a given frame. However, this does not necessarily
implies a loss in energy efficiency. As a matter of facts, the
simulation results we are going to present show that there is not a
scheme that outperforms the others in terms of both energy and
spectral efficiency, but which scheme is best depends on the maximum
throughput we want to achieve. In Fig. \ref{fig:through_energy} we
see that SA achieves a higher throughput with a lower energy
consumption with respect to the other schemes in the region
$\Phi<0.35$. In the region $\Phi>0.35$, instead, both NCDP and CRDSA
achieve a higher throughput with lower energy consumption with
respect to SA. NCDP and CRDSA behave almost in the same way in the
case of $2$ repetitions, achieving a maximum throughput of $0.5$ for
an average energy consumption of $2$. In the case of $3$ repetitions
NCDP achieves a maximum $\Phi$ of $0.7$, higher than CRDSA, for
which the peak value is $0.6$, for $\eta=3$. In the NCDP scheme with
a retransmission probability of $p=0.0453$ a peak throughput of
$0.8$ is achieved in correspondence of an average energy consumption
of $\eta=6.795$. For comparison, we also show the throughput-energy
curve for NCDP in case of $p=0.9961$, i.e., coefficients $\alpha$
are chosen uniformly in $GF(2^8)$. The high $p$ leads to a high
throughput, but also to a high energy consumption, with a minimum of
$\eta=149.415$. Moreover, we note that the gain with respect to the
scheme with $p=0.0453$ is negligible (about $5\%$), especially when
compared to the energy saving of about $95\%$ of this last one.
\section{Conclusions}\label{sec:conclusions}
We have proposed a new collision recovery scheme for symbol-synchronous
slotted ALOHA systems based on PHY layer NC over extended Galois
Fields. This allows to better exploit the diversity of the system,
leading to increased spectral efficiency and, depending on the
system load, to an increased energy efficiency. We have compared the
proposed scheme with two benchmark schemes in two different setups.
 One is a best-effort setup, in which the nodes do not receive any feedback from the receiver. In
the second setup feedback is allowed from the
receiver and an ARQ mechanism is assumed. In the second setup we have
evaluated jointly the spectral efficiency and the energy consumption
 of the proposed scheme and compared it with other
collision resolution schemes previously proposed in the literature.
 Once the PHY layer NC is applied to decode the collided
bursts, the receiver applies common matrix manipulation techniques
over finite fields, which results in a high-throughput scheme. The
increase in throughput coming from an increased number of
transmissions implies a higher energy consumption for a given
transmitter in a given frame. However, this does not necessarily
implies a loss in energy efficiency. We showed that NCDP achieves a
higher spectral efficiency with respect to the considered
benchmarks, while there is not a single scheme that outperforms the
others in terms of both energy and spectral efficiency, but the best
scheme depends on the maximum achievable throughput.

Furthermore, we carried out an analysis of several physical layer
issues related to multi-user PHY NC. We extended the analysis on and
proposed countermeasures against the effects of physical layer
impairments on the FER when applying PHY NC for a generic number of
colliding signals. In particular,
 we took into account frequency and
phase offsets at the transmitters which, up to our knowledge, have
been previously addressed only for the case of two colliding
signals. Finally, we showed the feasibility of channel estimation for PHY NC
in the presence of more than two colliding signals and studied
the effect of non perfect symbol synchronism on the decoder FER, proposing four different methods to compensate for such effect. Up to
our knowledge, this kind of analysis has been carried out only for
the case of two colliding signals and mainly in the context of two-way
relay communication.
\section*{Appendix}
Starting from the samples $r(t_l)$ the receiver $R$ wants to decode
the codeword $\mathbf{x}_s\triangleq\mathbf{x}_1\oplus \mathbf{x}_2 \oplus \ldots
\oplus \mathbf{x}_k$, where $\oplus$ denotes the bit-wise XOR. In
order to do this we must feed the decoder of $R$ with the vector
$\mathbf{L}^\oplus=\{L^\oplus(1),...,L^\oplus(N)\}$ of LLRs for
$\mathbf{x}_{s}$. We have:
\begin{eqnarray}\label{eqn:llr}
L^\oplus(l)&\triangleq&\ln
\left\{\frac{Pr\left[x_s(l)=1|r(t_l)\right]}{Pr\left[x_s(l)=0|r(t_l)\right]}\right\}\notag\\&=&
\ln\left\{\frac{Pr\left[r(t_l)|x_s(l)=1\right]}
{Pr\left[r(t_l)|x_s(l)=0\right]}\right\}.
\end{eqnarray}
The last equality follows from the symmetry of the XOR operator
provided that $x_j(l)$'s are independent and identically distributes
(i.i.d.) with $Pr[x_j(l)=1]=Pr[x_j(l)=0]=\frac{1}{2}$. Equation
(\ref{eqn:llr}) reduces to the calculation of the ratio of the
likelihood functions of $r(t_l)$ for the cases $x_s(l)=1$ and
$x_s(l)=0$. We indicate these functions as $f_{1}(r(t_l))$ and
$f_{0}(r(t_l))$ respectively. Functions $f_0(r(t_l))$ and
$f_1(r(t_l))$ are Gaussian mixtures:
\begin{eqnarray}\label{eqn:pdf1asymmetric}
f_1(r(t_l))=\frac{2^{-k}}{\sqrt{2\pi
N_0}}\sum_{i=1}^{\lfloor\frac{k+1}{2}\rfloor} \sum_{m=1}^{{k \choose
2i-1}} e^{-
\frac{\left|r(t_l)-\mathbf{d}^o(2i-1,m)^T\mathbf{h}(t_l)\right|^2}{2N_0}
},
\end{eqnarray}
$\mathbf{h}(t_l)$ being a column vector containing the channel
coefficients of the $k$ transmitters at time $t_l$ (which change at
each sample due to frequency offsets), while $\mathbf{d}^{o}(2i-1,m)$ is a column vector containing one
(the m-th) of the ${k \choose 2i-1}$ possible permutations over $k$
symbols (without repetitions) of an odd number ($2i-1$) of symbols
with value ``$+1$". As for the case with $x_s=0$ we have:
\begin{eqnarray}\label{eqn:pdf0asymmetric}
f_0(r(t_l))=\frac{2^{-k}}{\sqrt{2\pi
N_0}}\sum_{i=1}^{\lfloor\frac{k+1}{2}\rfloor} \sum_{m=1}^{{k \choose
2i}} e^{-
\frac{\left|r(t_l)-\mathbf{d}^e(2i,m)^T\mathbf{h}(t_l)\right|^2}{2N_0}
},
\end{eqnarray}
where $\mathbf{d}^{e}(2i,m)$ is a column vector containing one (the
m-th) of the ${k \choose 2i}$ possible permutations over $k$ symbols
(without repetitions) of an even number ($2i$) of symbols with value
``$+1$". Finally using (\ref{eqn:pdf1asymmetric}) and
(\ref{eqn:pdf0asymmetric}) in (\ref{eqn:llr}) we find the following
expression for the LLR:
\begin{eqnarray}\label{eqn:llrasymmetric_app}
L^{\oplus}(l)=\ln \left\{
\frac{\sum_{i=1}^{\lfloor\frac{k+1}{2}\rfloor} \sum_{m=1}^{{k
\choose 2i-1}} e^{-
\frac{\left|r(t_l)-\mathbf{d}^o(2i-1,m)^T\mathbf{h}(t_l)\right|^2}{2N_0}
}}{\sum_{i=1}^{\lfloor\frac{k+1}{2}\rfloor} \sum_{m=1}^{{k \choose
2i}} e^{-
\frac{\left|r(t_l)-\mathbf{d}^e(2i,m)^T\mathbf{h}(t_l)\right|^2}{2N_0}
}}\right\}.
\end{eqnarray}
\bibliographystyle{IEEEbib}
\bibliography{cocco12_JSAC}

\begin{thebibliography}{10}

\bibitem{Choudhury_83_DSA}
G.~Choudhury and S.~Rappaport,
\newblock ``Diversity {ALOHA} - a random access scheme for satellite
  communications,''
\newblock {\em IEEE Trans. on Comm.}, vol. 31, no. 3, pp. 450--457, Mar. 1983.

\bibitem{casini07_CRDSA_aloha}
E.~Casini, R.~De~Gaudenzi, and O.~d.~R.~Herrero,
\newblock ``Contention resolution diversity slotted {ALOHA} ({CRDSA}): An
  enhanced random access scheme for satellite access packet networks,''
\newblock {\em IEEE Trans. on Wireless Comm.}, vol. 6, no. 4, pp. 1408--1419,
  Apr. 2007.

\bibitem{bui_2010_phy_nc_aloha}
H.~C. Bui, J.~Lacan, and M.-L. Boucheret,
\newblock ``{NCSA}: A new protocol for random multiple access based on physical
  layer network coding,'' http://arxiv.org/abs/1009.4773, 2010.

\bibitem{Liva_11_CRDSA}
G.~Liva,
\newblock ``Graph-based analysis and optimization of contention resolution
  diversity slotted {ALOHA},''
\newblock {\em IEEE Trans. on Comm.}, vol. 59, no. 2, pp. 477--487, Feb. 2011.

\bibitem{zhang2006_phy_nc}
S.~Zhang, S.~Liew, and P.~Lam,
\newblock ``Physical layer network coding,''
\newblock in {\em ACM MOBICOM}, Los Angeles (CA), U.S.A., Sep. 2006.

\bibitem{rossetto09_design_asyn_PHY_NC}
F.~Rossetto and M.~Zorzi,
\newblock ``On the design of practical asynchronous physical layer network
  coding,''
\newblock in {\em IEEE Workshop on Signal Proc. Advances in Wireless Comm.},
  Perugia, Italy, June 2009.

\bibitem{louie10_practica_PHY_NC}
R.~H.~Y. Louie, Y.~Li, and B.~Vucetic,
\newblock ``Practical physical layer network coding for two-way relay channels:
  Performance analysis and comparison,''
\newblock {\em IEEE Trans. on Wireless Comm.}, vol. 9, no. 2, pp. 764--777,
  Feb. 2010.

\bibitem{sorensen09_PHY_NC_FSK}
J.~H. Sorensen, R.~Krigslund, P.~Popovski, T.~Akino, and T.~Larsen,
\newblock ``Physical layer network coding for {FSK} systems,''
\newblock {\em IEEE Comm. Letters}, vol. 13, no. 8, Aug. 2009.

\bibitem{rossetto_asms_2010}
F.~Rossetto,
\newblock ``A comparison of different physical layer network coding techniques
  for the satellite environment,''
\newblock in {\em {Advanced Satellite Multimedia Systems Conf. (ASMS)}},
  Cagliari, Italy, Sep. 2010.

\bibitem{durvy07_reliable_broadcast}
M.~Durvy, C.~Fragouli, and P.~Thiran,
\newblock ``Towards reliable broadcasting using {ACKs},''
\newblock in {\em IEEE Int'l Symp. on Info. Theo. (ISIT)}, Nice, France, June
  2007.

\bibitem{foh10_collision_codes}
C.~H. Foh, J.~Cai, and J.~Qureshi,
\newblock ``Collision codes: Decoding superimposed {BPSK} modulated wireless
  transmissions,''
\newblock in {\em IEEE Consumer Comm. and Networking Conf.}, Las Vegas (NV),
  U.S.A., Jan. 2010.

\bibitem{nazer11_reliable_phy_nc}
B.~Nazer and M.~Gastpar,
\newblock ``Reliable physical layer network coding,''
\newblock {\em Proceedings of the IEEE}, vol. 99, no. 3, pp. 438--460, Mar.
  2011.

\bibitem{cocco11_mu_phy_nc_aloha}
G.~Cocco, C.~Ibars, D.~G\"{u}nd\"{u}z, and O.~d.~R.~Herrero,
\newblock ``Collision resolution in slotted {ALOHA} with multi-user
  physical-layer network coding,''
\newblock in {\em IEEE Vehicular Technology Conf. (VTC Spring)}, Budapest,
  Hungary, May 2011.

\bibitem{maduike09_phy_nc_offsets}
D.~Maduike, H.~D. Pfister, and A.~Sprintson,
\newblock ``Design and implementation of physical-layer network-coding
  protocols,''
\newblock in {\em Asilomar Conference on Signals, Systems and Computers},
  Pacific Grove (CA), U.S.A., Nov. 2009.

\bibitem{lulu12_asynchronous_phy_nc}
L.~Lu and S.~C. Liew,
\newblock ``Asynchronous physical-layer network coding,''
\newblock {\em IEEE Trans. on Wireless Comm.}, vol. 11, no. 2, pp. 819--831,
  Feb. 2012.

\bibitem{jain11_param_estim_phy_nc}
M.~Jain, S.~L. Miller, and A.~Sprintson,
\newblock ``Parameter estimation and tracking in physical layer network
  coding,''
\newblock in {\em IEEE Global Telecomm. Conf.}, Houston (TX), U.S.A., Dec.
  2011.

\bibitem{koetter01_algebraic_NC}
R.~Koetter and M.~Medard,
\newblock ``An algebraic approach to network coding,''
\newblock in {\em IEEE Int'l Symp. on Info. Theo.}, Washington, D.C., U.S.A.,
  June 2001.

\bibitem{trullols11_exact_NC}
O.~Trullols-Cruces, J.~M. Barcelo-Ordinas, and M.~Fiore,
\newblock ``Exact decoding probability under random linear network-coding,''
\newblock {\em IEEE Comm. Letters}, vol. 15, no. 1, pp. 67--69, Jan. 2011.

\bibitem{blomer97_rank_sparse_matrix}
J.~Bl\"{o}mer, R.~Karp, and E.~Welzl,
\newblock ``The rank of sparce random matrices over finite fiels,''
\newblock {\em Random Structures and Algorithms}, vol. 10, no. 4, pp. 407--419,
  July 1997.

\bibitem{feder88_param_estim_EM}
M.~Feder and E.~Weinstein,
\newblock ``Parameter estimation of superimposed signals using the {EM}
  algorithm,''
\newblock {\em IEEE Trans. on Acoustics, Speech and Signal Processing}, vol.
  36, no. 4, pp. 477--489, Apr. 1988.

\bibitem{nee94_cdma_synchro}
R.~D. J.~Van Nee,
\newblock ``Timing aspects of synchronous {CDMA},''
\newblock in {\em IEEE Int'l Symp. on Personal, Indoor and Mobile Radio Comm.},
  The Hague, The Netherlands, Sep. 1994.

\bibitem{Kleinrock75_performance}
L.~Kleinrock and S.~Lam,
\newblock ``Packet switching in a multiaccess broadcast channel: Performance
  evaluation,''
\newblock {\em IEEE Trans. on Comm.}, vol. 23, no. 4, pp. 410--423, Apr 1975.

\end{thebibliography}
\end{document}